\documentclass{article}

\usepackage[utf8]{inputenc}
\usepackage[T1]{fontenc}

\PassOptionsToPackage{table, dvipsnames}{xcolor}
\usepackage{xcolor}

\PassOptionsToPackage{colorlinks=true, citecolor=blue, linkcolor=blue, urlcolor=black}{hyperref}

\usepackage{arxiv}
\usepackage[utf8]{inputenc}
\usepackage[T1]{fontenc}
\usepackage{url}
\usepackage{booktabs}
\usepackage{nicefrac}
\usepackage{microtype}
\usepackage{lipsum}
\usepackage{graphicx}
\usepackage{natbib}
\usepackage{doi}
\usepackage{algorithm}
\usepackage{multirow}
\usepackage{multicol}
\usepackage{amsmath,amssymb,amsfonts}
\usepackage{amsthm}
\usepackage{mathrsfs}
\usepackage{float}
\usepackage{bm}
\usepackage{caption}
\usepackage{hyperref}
\usepackage{algpseudocode}

\title{Propagating the prior from far to near offset: A self-supervised diffusion framework for progressively recovering near-offsets of towed-streamer data
}

\author{ \href{https://orcid.org/0000-0001-8868-7967}{\includegraphics[scale=0.06]{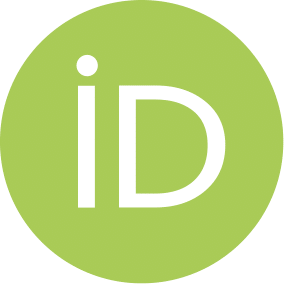}\hspace{1mm}Shijun~Cheng}\\
	Division of Physical Science and Engineering\\
	King Abdullah University of Science and Technology\\
	Thuwal 23955-6900, Saudi Arabia \\
	\texttt{sjcheng.academic@gmail.com} \\
        \And
	\href{https://orcid.org/0000-0002-5950-5201}{\includegraphics[scale=0.06]{orcid.png}\hspace{1mm}Tariq~Alkhalifah} \\
	Division of Physical Science and Engineering\\
	King Abdullah University of Science and Technology\\
	Thuwal 23955-6900, Saudi Arabia \\
	\texttt{tariq.alkhalifah@kaust.edu.sa} \\
}

\renewcommand{\shorttitle}{A self-supervised diffusion framework for near-offset reconstruction}

\hypersetup{
pdftitle={A template for the arxiv style},
pdfsubject={q-bio.NC, q-bio.QM},
pdfauthor={David S.~Hippocampus, Elias D.~Striatum},
pdfkeywords={First keyword, Second keyword, More},
}

\begin{document}
\maketitle

\begin{abstract}
In marine towed-streamer seismic acquisition, the nearest hydrophone is often two hundred meter away from the source resulting in missing near-offset traces, which degrades critical processing workflows such as surface-related multiple elimination, velocity analysis, and full-waveform inversion. Existing reconstruction methods, like transform-domain interpolation, often produce kinematic inconsistencies and amplitude distortions, while supervised deep learning approaches require complete ground-truth near-offset data that are unavailable in realistic acquisition scenarios. To address these limitations, we propose a self-supervised diffusion-based framework that reconstructs missing near-offset traces without requiring near-offset reference data. Our method leverages overlapping patch extraction with single-trace shifts from the available far-offset section to train a conditional diffusion model, which learns offset-dependent statistical patterns governing event curvature, amplitude variation, and wavelet characteristics. At inference, we perform trace-by-trace recursive extrapolation from the nearest recorded offset toward zero offset, progressively propagating learned prior information from far to near offsets. The generative formulation further provides uncertainty estimates via ensemble sampling, quantifying prediction confidence where validation data are absent. Controlled validation experiments on synthetic and field datasets show substantial performance gains over conventional parabolic Radon transform baselines. Operational deployment on actual near-offset gaps demonstrates practical viability where ground-truth validation is impossible. Notably, the reconstructed waveforms preserve realistic amplitude-versus-offset trends despite training exclusively on far-offset observations, and uncertainty maps accurately identify challenging extrapolation regions, offering practitioners actionable quality indicators for downstream processing decisions.
\end{abstract}

\section{\textbf{Introduction}}
In marine seismic surveys, towed-streamer acquisition has become a popular method for recording subsurface reflection data, thanks to its operational efficiency, extensive areal coverage, and high-density multichannel capability \citep{widmaier2019redefining}. However, the required fixed lateral separation between the airgun source and the first hydrophone group unavoidably omits the recording of true near-offset traces. This loss of near-offset data undermines key processing workflows. For example, surface-related multiple elimination (SRME) \citep{verschuur1992adaptive} fundamentally depends on near-offset recordings to assemble a source-over-receiver acquisition geometry for accurate multidimensional convolution, and its efficacy is severely degraded in their absence \citep{cheng2025self}. Likewise, first-break tomography, amplitude-versus-offset analysis, and full-waveform inversion all could benefit from small-angle, near zero offset, reflection information to constrain velocity models and increase velocity  resolution \citep{swan2001velocities, alkhalifah2014scattering, salaun2020capturing, jimenez2022downward}. Therefore, the reconstruction of missing near-offset data is essential to restore processing integrity and enhance the accuracy of subsequent seismic imaging and interpretation.

To mitigate the near-offset gap at the acquisition stage, modern surveys have adopted source-over-streamer geometries such as TopSeis, wherein airgun arrays are towed directly above the streamer, thereby capturing true zero- and near-offset traces without additional hardware. When acquisition modifications are infeasible or incomplete, processing-based reconstructions are applied. The most elementary workflow begins with conventional NMO flattening of primaries, copies the nearest recorded trace to fill each absent near-offset position, and then reverses the NMO stretch to restore the true geometry \citep{curry2010interpolation, cheng2025self}. More advanced prediction-error filter (PEF) methods estimate nonstationary filters from mid- and far-offset data to extrapolate the missing near-offset samples while preserving amplitude \citep{curry2010interpolation}. Transform-domain interpolation, such as F-X predictive filtering, which exploits local linearity in the frequency-space domain, and hyperbolic Radon mappings that coherently remap energy into the zero-offset region-offers high-fidelity gap filling with accurate kinematics \citep{kabir1995restoration, sacchi1997recovery, trad2002accurate, xu2018radon}. Finally, multiples-based reconstruction leverages SRME to synthesize pseudoprimary near-offset traces by convolving recorded primaries and multiples, effectively generating a virtual source-receiver survey \citep{van2009estimating, van2009estimation, lopez2015closed, zhang2019integration}. In practice, hybrid workflows that couple nearest-trace or PEF with transform-domain and multiples-based operators are routinely employed to optimize computational efficiency, amplitude consistency, and kinematic accuracy before proceeding to velocity analysis and imaging.

Recent advances in deep learning have introduced data-driven alternatives for seismic trace reconstruction, leveraging deep neural networks to learn complex mappings from incomplete to complete wavefields \citep{jia2017can, wang2019deep, harsuko2022storseismic, cheng2023seismic, mousavi2024applications}. U-Net architectures, originally designed for image segmentation \citep{ronneberger2015u}, have been adapted to interpolate missing seismic traces by exploiting hierarchical feature extraction and spatial context across multiple scales \citep{park2019reconstruction, chai2020deep, park2021method, fang2021seismic, he2021seismic, cheng2024meta}. Long short-term memory networks capture temporal dependencies in seismic sequences, enabling coherent prediction of missing samples along offset or time axes \citep{yoon2020seismic}. Generative adversarial networks (GANs) have demonstrated the ability to synthesize realistic seismic data by training generator-discriminator pairs, thereby recovering missing traces with enhanced high-frequency content and structural fidelity \citep{oliveira2018interpolating, chang2020seismic, kaur2021seismic, wei2021reconstruction}. In the context of towed-streamer near-offset reconstruction, supervised deep learning methods have shown promise by training on paired datasets where complete synthetic sections are artificially degraded to simulate acquisition gaps \citep{qu2021training, huff2024near}. These approaches typically achieve superior reconstruction quality compared to conventional signal-processing techniques, particularly in handling complex subsurface structures and noisy field conditions.

Despite their success, supervised learning methods face a fundamental generalization challenge: models trained on synthetic datasets often fail to adapt to the field data \citep{alkhalifah2022mlreal}. The scarcity of labeled training pairs further constrains the practical deployment of these approaches, as ground-truth complete data are rarely available in actual towed-streamer acquisitions. To address this limitation, self-supervised learning (SSL) paradigms have emerged as a promising alternative, enabling models to learn reconstruction mappings directly from incomplete observed data without requiring complete references. To the best of our knowledge, only two SSL methods have been reported for near-offset reconstruction in towed-streamer data. \cite{wang2022missing} proposed a single-stage self-supervised framework that reconstructs regularly missing shots and near-offset data by exploiting the spatial reciprocity of Green’s functions and by generating pseudo near-offset labels through a rotation–truncation procedure. Their method learns directly from field data without requiring synthetic pretraining, enabling simultaneous recovery of missing shots and near-offset gaps. However, this strategy requires the near-offset dip to become approximately zero after rotation, and selecting an appropriate rotation angle is not trivial. An inaccurate rotation can distort the reflection geometry and substantially degrade reconstruction performance. In contrast, \cite{park2024near} focused specifically on near-offset extrapolation and introduced a two-stage self-supervised learning paradigm. Their approach first pretrains a U-Net3+ (a full-scale connected UNet) model on a diverse synthetic dataset with artificially created near-offset gaps to learn generic near-offset representations, and subsequently adapts the model to field data through self-supervised transfer learning, where pseudo-labels are generated by masking and restoring the nearest recorded offsets. During inference, the adapted model extrapolates the actual near-offset gap using patch-based prediction. Nonetheless, the success of this two-stage strategy depends strongly on the diversity and representativeness of the synthetic data used for pretraining, and the patch-based inference can interrupt long-range structural continuity, resulting in stitching artifacts when extrapolating the near-offset region.

In this work, we introduce a self-supervised generative diffusion model (SSGDM) for near-offset reconstruction in towed-streamer data that addresses the aforementioned limitations. The key innovation of our approach lies in a recursive inference strategy: we train a conditional diffusion model on the available far-offset data and then progressively extrapolate trace-by-trace from the nearest recorded offset toward the missing near-offset region. During training, we exploit the spatial redundancy in the recorded far-offset section by extracting pairs of overlapping patches that are laterally shifted by a single trace. The patch closer to zero offset serves as the target distribution $x_0$, while the farther patch provides the conditioning context $y$, enabling the model to learn the conditional distribution $p(x_0|y)$. This training paradigm allows the diffusion model to capture the statistical patterns governing event curvature, amplitude variation, and wavelet characteristics as they evolve across from far to near offset. At inference time, we initialize the reconstruction from the nearest available trace and iteratively predict each successive missing trace by conditioning on the previously reconstructed (or recorded) offset, thereby propagating the learned prior information from far offsets toward near offsets in a physically consistent manner. Crucially, this recursive formulation eliminates the need for complete ground-truth data while naturally accommodating the directional extrapolation required for filling the near-offset gap. Furthermore, the probabilistic nature of diffusion models enables uncertainty quantification through ensemble sampling: by generating multiple realizations of the missing near-offset section, we can assess reconstruction confidence and identify regions where extrapolation uncertainty is elevated. We validate the proposed method on synthetic examples and two field marine datasets, demonstrating superior reconstruction fidelity and meaningful uncertainty estimates.

\section{\textbf{Methodology}}

\subsection{Conditional diffusion models for seismic interpolation}

Denoising diffusion probabilistic models (DDPMs) are a class of generative models that synthesize data samples by learning to reverse a gradual noising process \citep{ho2020denoising, nichol2021improved}. The forward diffusion process progressively corrupts a clean data sample $x_0 \sim q(x_0)$ by adding Gaussian noise over $T$ time steps according to a predefined variance schedule $\{\beta_t\}_{t=1}^T$:
\begin{equation}\label{eq1}
q(x_t|x_{t-1}) = \mathcal{N}(x_t; \sqrt{1-\beta_t}x_{t-1}, \beta_t\mathbf{I}),
\end{equation}
where $x_t$ denotes the noisy sample at time step $t$. Using a reparameterization trick, the forward process can be directly sampled at any step $t$ as:
\begin{equation}\label{eq2}
x_t = \sqrt{\bar{\alpha}_t}x_0 + \sqrt{1-\bar{\alpha}_t}\epsilon, \quad \epsilon \sim \mathcal{N}(0, \mathbf{I}),
\end{equation}
where $\alpha_t = 1 - \beta_t$ and $\bar{\alpha}_t = \prod_{s=1}^t \alpha_s$. As $t$ increases, $x_t$ gradually loses structure and converges to a pure Gaussian noise when $t \to T$.

The reverse diffusion process learns to iteratively denoise $x_T \sim \mathcal{N}(0, \mathbf{I})$ back to a sample from the data distribution $q(x_0)$. This can be achieved by training a neural network (NN) to predict either the added noise $\epsilon$ or the clean data $x_0$ directly. Recent studies in seismic processing have demonstrated that directly predicting $x_0$ yields superior reconstruction quality compared to noise prediction \citep{cheng2025gsfm}. Following this finding, we adopt the $x_0$-prediction formulation, where an NN $f_\theta(x_t, t)$ is trained to predict the clean data from the noisy observation:
\begin{equation}\label{eq3}
\mathcal{L}_{\text{simple}} = \mathbb{E}_{t, x_0, \epsilon}\left[\|x_0 - f_\theta(x_t, t)\|^2\right],
\end{equation}
where $x_t$ is obtained by adding noise to $x_0$ according to Equation~\ref{eq2}. Given the predicted $\hat{x}_0 = f_\theta(x_t, t)$, the reverse transition can be formulated as:
\begin{equation}\label{eq4}
p_\theta(x_{t-1}|x_t) = \mathcal{N}(x_{t-1}; \mu_\theta(x_t, t), \sigma_t^2\mathbf{I}),
\end{equation}
where the predicted mean is computed as:
\begin{equation}\label{eq5}
\mu_\theta(x_t, t) = \frac{\sqrt{\bar{\alpha}_{t-1}}\beta_t}{1-\bar{\alpha}_t}\hat{x}_0 + \frac{\sqrt{\alpha_t}(1-\bar{\alpha}_{t-1})}{1-\bar{\alpha}_t}x_t.
\end{equation}

In the context of seismic data interpolation, conditional diffusion models have been employed to reconstruct missing traces by conditioning the generative process on available observations \citep{durall2023deep, liu2024generative, wang2025self}. Specifically, let $x_0$ denote the complete target data (including both observed and missing traces), and $y$ denote the observed partial data (with missing traces set to zero or masked). The conditional diffusion model learns to generate $x_0$ conditioned on $y$ by training a network $f_\theta(x_t, y, t)$ to predict the clean complete data from the noisy version $x_t$ and the conditioning context $y$:
\begin{equation}\label{eq6}
\mathcal{L}_{\text{cond}} = \mathbb{E}_{t, x_0, y, \epsilon}\left[\|x_0 - f_\theta(x_t, y, t)\|^2\right].
\end{equation}
During training, $y$ is typically concatenated with $x_t$ along the channel dimension and fed into the denoising network, enabling the model to learn the conditional distribution $p(x_0|y)$ \citep{liu2024generative}. At inference time, the denoising diffusion model iteratively refines random noise into a reconstructed complete section that is consistent with the observed traces $y$.

However, this supervised training paradigm requires access to complete ground-truth data $x_0$ (i.e., seismic sections with no missing traces) paired with their artificially degraded versions $y$ (where traces are synthetically removed to simulate acquisition gaps). In practice, such complete near-offset data are inherently unavailable due to the physical source-receiver separation, making supervised training infeasible. Moreover, models trained on synthetic data based degradation patterns often fail to generalize to the specific acquisition geometries and subsurface complexities encountered in field data. Therefore, developing an SSL framework that can learn reconstruction mappings directly from incomplete observed data remains a critical challenge for near-offset reconstruction.

\subsection{Self-supervised conditional diffusion model for near-offset reconstruction}

Our training strategy leverages the spatial redundancy in the recorded far-offset section to construct self-supervised training pairs without requiring complete near-offset ground truth. Let $\mathbf{D} \in \mathbb{R}^{N_s \times N_t \times N_r}$ denote the complete desired marine seismic dataset corresponding to a towed-streamer geometry, where $N_s$, $N_t$, and $N_r$ represent the number of shots, time samples, and receivers, respectively. In this complete scenario, each shot gather $\mathbf{D}_i \in \mathbb{R}^{N_t \times N_r}$ ($i = 1, \ldots, N_s$) contains recordings from $N_r$ receivers, ordered by increasing offset from the source. Due to the fixed source-receiver separation in towed-streamer acquisition, we assume that the first $N_m$ near-offset traces are missing, and only the far-offset traces $\{N_m+1, N_m+2, \ldots, N_r\}$ are recorded. Let $\mathbf{D}_{\text{obs}} \in \mathbb{R}^{N_s \times N_t \times (N_r - N_m)}$ denote the actual recorded observed data containing only the recorded far-offset traces, and $\mathbf{D}_{\text{miss}} \in \mathbb{R}^{N_s \times N_t \times N_m}$ denote the missing near-offset region that we aim to reconstruct.

For each training iteration, we randomly select a shot gather from $\mathbf{D}_{\text{obs}}$ and extract two overlapping patches of width $W$ traces. Specifically, let $x_0 \in \mathbb{R}^{N_t \times W}$ denote the target patch comprising traces from indices $[r_i, r_i+W-1]$, and $y \in \mathbb{R}^{N_t \times W}$ denote the conditioning patch comprising traces from indices $[r_i+1, r_i+W]$. Thus, the two patches are laterally shifted by a single trace, with $x_0$ positioned one trace closer to zero-offset than $y$:
\begin{equation}\label{eq7}
x_0 = \mathbf{D}_{\text{obs}}(j, :, r_i:r_i+W-1), \quad y = \mathbf{D}_{\text{obs}}(j, :, r_i+1:r_i+W),
\end{equation}
where $j$ is a randomly sampled shot index from $\{1, \ldots, N_s\}$, and $r_i$ is uniformly sampled from the range $\{N_m + 1, \ldots, N_r - W\}$ to ensure that both patches lie entirely within the recorded far-offset aperture.

Given the training pair $(x_0, y)$, we synthesize a noisy version $x_t$ by applying the forward diffusion process (Equation~\ref{eq2}) to $x_0$ at a randomly sampled time step $t \sim \text{Uniform}(1, T)$:
\begin{equation}\label{eq8}
x_t = \sqrt{\bar{\alpha}_t}x_0 + \sqrt{1-\bar{\alpha}_t}\epsilon, \quad \epsilon \sim \mathcal{N}(0, \mathbf{I}).
\end{equation}
The conditional diffusion model $f_\theta(x_t, y, t)$ is then trained to predict the clean target $x_0$ from the noisy input $x_t$ and the conditioning context $y$. The conditioning patch $y$ is concatenated with $x_t$ along the channel dimension, forming an input of size $\mathbb{R}^{2 \times N_t \times W}$, which is then processed by a U-Net architecture based neural network with 2D convolutional layers. The training loss is formulated as the mean squared error (MSE) between the true target and the predicted reconstruction:
\begin{equation}\label{eq9}
\mathcal{L} = \mathbb{E}_{j, r_i, t, \epsilon}\left[\|x_0 - f_\theta(x_t, y, t)\|_2^2\right].
\end{equation}
By minimizing this loss over the entire training dataset, given by windows of the available recorded data, the model learns to capture the statistical dependencies governing how seismic events evolve as a function of offset. Crucially, this training paradigm does not require any data from the missing near-offset region. Instead, it solely relies on the self-consistency of the recorded far-offset traces. The overlapping patch design, with a single-trace shift, explicitly trains the model to perform the exact extrapolation task required during inference.

\subsection{Recursive Inference with DDIM Sampling}
At inference time, we reconstruct the missing near-offset traces in a sequential, trace-by-trace manner, starting from the nearest recorded trace and progressively extrapolating toward zero offset. Let $r_{\text{start}} = N_m + 1$ denote the index of the first recorded trace (i.e., the trace closest to the missing region), and let $r_k = N_m, N_m - 1, \ldots, 1$ enumerate the missing trace indices in descending order.

To reconstruct the trace at index $r_k$, we construct the conditioning context $y_k \in \mathbb{R}^{N_t \times W}$ by assembling $W$ consecutive traces starting from index $r_k + 1$:
\begin{equation}\label{eq10}
y_k = [\hat{x}_{r_k+1}, \hat{x}_{r_k+2}, \ldots, \hat{x}_{r_k+W}],
\end{equation}
where $\hat{x}_{r_k+i}$ denotes the reconstructed trace at index $r_k+i$ if $r_k+i < r_{\text{start}}$, or the recorded trace if $r_k+i \geq r_{\text{start}}$.

To accelerate the sampling process, we employ the denoising diffusion implicit model (DDIM) \citep{song2020denoising}, which enables deterministic and faster sampling by directly computing large-step transitions in the reverse process. DDIM selects a uniformly spaced subsequence of $S$ time steps $\{t_1, t_2, \ldots, t_S\}$ from the full training schedule $\{0, 1, \ldots, T\}$, where $t_S = T$ corresponds to pure Gaussian noise and $t_0 = 0$ corresponds to clean data. With $S \ll T$ (e.g., $S=50$ for $T=1000$), DDIM reduces the number of function evaluations from $T$ to $S$ while maintaining high-quality generation.

We initialize sampling from pure Gaussian noise $x_T \sim \mathcal{N}(\mathbf{0}, \mathbf{I})$ with shape $\mathbb{R}^{N_t \times W}$, representing the maximum noise level at diffusion time $t_S = T$. This noise sample is concatenated with the conditioning context $y_k$ along the channel dimension to form the network input of shape $\mathbb{R}^{2 \times N_t \times W}$. At each reverse step $s = S, S-1, \ldots, 1$, given the noisy sample $x_{t_s}$ at diffusion time $t_s$, we predict the clean data $\hat{x}_0 = f_\theta(x_{t_s}, y_k, t_s)$ and update to the next diffusion time $t_{s-1}$ using:
\begin{equation}\label{eq11}
x_{t_{s-1}} = \sqrt{\bar{\alpha}_{t_{s-1}}}\hat{x}_0 + \sqrt{1 - \bar{\alpha}_{t_{s-1}}}\cdot\frac{x_{t_s} - \sqrt{\bar{\alpha}_{t_s}}\hat{x}_0}{\sqrt{1-\bar{\alpha}_{t_s}}}.
\end{equation}

After $S$ reverse diffusion steps, we obtain the final reconstruction $\hat{x}_0 \in \mathbb{R}^{N_t \times W}$, which has the same spatial extent as the conditioning context $y_k$ but is shifted by one trace toward zero offset. Specifically, $\hat{x}_0$ comprises traces from indices $[r_k, r_k+1, \ldots, r_k+W-1]$, effectively discarding the farthest trace $\hat{x}_{r_k+W}$ from $y_k$ and generating the new near-offset trace $\hat{x}_{r_k}$. We extract the first trace from the reconstructed patch and incorporate it into the conditioning context for the next iteration, i.e., $y_k \leftarrow \hat{x}_0 = [\hat{x}_{r_k}, \hat{x}_{r_k+1}, \ldots, \hat{x}_{r_k+W-1}]$. This recursive process continues until all missing traces are reconstructed up to zero offset. The complete recursive inference procedure is summarized in Algorithm~\ref{alg:recursive_inference}.

\begin{algorithm}[t]
\caption{Recursive near-offset reconstruction via DDIM sampling}
\label{alg:recursive_inference}
\begin{algorithmic}[1]
\Require Observed far-offset data $\mathbf{D}_{\text{obs}}$, trained model $f_\theta$, patch width $W$, DDIM steps $S$, number of missing traces $N_m$
\Ensure Reconstructed near-offset section $\{\hat{x}_{1}, \hat{x}_{2}, \ldots, \hat{x}_{N_m}\}$
\State Let $\{t_1, t_2, \ldots, t_S\}$ be uniformly spaced DDIM time steps with $t_S = T$ and $t_0 = 0$
\State Initialize $r_{\text{start}} \gets N_m + 1$ \Comment{First recorded trace index}
\State Initialize reconstructed traces: $\hat{x}_i \gets \mathbf{D}_{\text{obs}}(:, i)$ for $i \geq r_{\text{start}}$ \Comment{Copy recorded traces}
\For{$k = N_m, N_m-1, \ldots, 1$} \Comment{Iterate from far to near offset}
    \State Construct conditioning context: $y_k \gets [\hat{x}_{k+1}, \hat{x}_{k+2}, \ldots, \hat{x}_{k+W}]$
    \State Initialize random noise: $x_T \sim \mathcal{N}(\mathbf{0}, \mathbf{I})$ with shape $\mathbb{R}^{N_t \times W}$ \Comment{$t_S = T$ (max noise)}
    \For{$s = S, S-1, \ldots, 1$} \Comment{DDIM reverse diffusion}
        \State Predict clean data: $\hat{x}_0 \gets f_\theta(x_{t_s}, y_k, t_s)$
        \State Update: $x_{t_{s-1}} \gets \sqrt{\bar{\alpha}_{t_{s-1}}}\hat{x}_0 + \sqrt{1 - \bar{\alpha}_{t_{s-1}}}\cdot\dfrac{x_{t_s} - \sqrt{\bar{\alpha}_{t_s}}\hat{x}_0}{\sqrt{1-\bar{\alpha}_{t_s}}}$
    \EndFor
    \State Extract reconstructed trace: $\hat{x}_k \gets \hat{x}_0(:, 1)$ \Comment{First trace of output}
\EndFor
\State \Return $\{\hat{x}_{1}, \hat{x}_{2}, \ldots, \hat{x}_{N_m}\}$
\end{algorithmic}
\end{algorithm}

This recursive formulation ensures that each newly reconstructed trace is conditioned on the most recently available offset information, thereby maintaining kinematic and amplitude consistency across the reconstructed near-offset section. Crucially, the single-trace shift between the predicted patch $\hat{x}_0$ and the conditioning context $y_k$ mirrors the training strategy, where the target patch $x_0$ and conditioning patch $y$ are also shifted by one trace, ensuring perfect alignment between training and inference procedures. Moreover, this single-channel incremental strategy facilitates smooth and gradual extrapolation, as seismic amplitudes, traveltimes, and waveforms vary continuously between adjacent offsets rather than exhibiting abrupt changes, thereby reducing the risk of accumulated reconstruction errors during recursive propagation.

\subsection{Uncertainty quantification via ensemble sampling}

A key advantage of the diffusion-based framework is its ability to naturally quantify reconstruction uncertainty. Unlike deterministic NNs that produce a single-point estimate, diffusion models generate samples from the learned conditional distribution $p(x_0|y)$, allowing us to assess the variability and confidence in the reconstruction.

For each missing trace, we perform $M$ independent DDIM sampling runs with different random noise initializations $x_{t_S}^{(m)} \sim \mathcal{N}(0, \mathbf{I})$, $m = 1, \ldots, M$, generating an ensemble of realizations $\{\hat{x}_0^{(m)}\}_{m=1}^M$. The mean reconstruction is computed as:
\begin{equation}\label{eq12}
\bar{x}_0 = \frac{1}{M}\sum_{m=1}^M \hat{x}_0^{(m)},
\end{equation}
and the pointwise standard deviation provides a pixel-wise uncertainty map:
\begin{equation}\label{eq13}
\sigma(x_0) = \sqrt{\frac{1}{M-1}\sum_{m=1}^M \left(\hat{x}_0^{(m)} - \bar{x}_0\right)^2}.
\end{equation}
High standard deviation values highlight regions where the model exhibits greater predictive uncertainty, typically corresponding to areas that involve challenging extrapolation from the available conditioning data. Such uncertainty estimates are particularly useful for quality control and risk assessment in seismic processing workflows, as they pinpoint traces or events where the reconstruction may be less reliable based on the available data and may therefore require additional estimate or alternative processing strategies.

\subsection{Network Architecture}
The conditional diffusion model employs a U-Net architecture adapted from improved DDPM \citep{nichol2021improved}. The network accepts a 2-channel input formed by concatenating the noisy target patch $x_t$ and conditioning patch $y$, and outputs a single-channel prediction $\hat{x}_0$. The U-Net comprises an encoder-decoder structure with five resolution scales. The encoder progressively downsamples the input through strided convolutions, applying channel multipliers $(1, 2, 4, 8, 16)$ to a base dimension of 64 channels, yielding feature maps of 64, 128, 256, 512, and 1024 channels at successive levels. Each scale contains two residual blocks with adaptive group normalization for time step conditioning via learned scale-shift parameters. The decoder mirrors the encoder structure, upsampling features through transposed convolutions. The symmetric skip connections directly concatenate encoder features with corresponding decoder features at each scale, enabling the network to recover fine-grained spatial details lost during downsampling. Self-attention layers with 4 heads are incorporated at the third and fourth scales (downsampling factors of 8 and 16) in both the encoder and decoder paths to capture long-range spatial correlations.
\section{\textbf{Controlled validation experiments}}
To evaluate the performance of the proposed self-supervised framework under controlled conditions, we conduct comprehensive experiments on synthetic and field marine seismic datasets where ground-truth near-offset traces are available for quantitative assessment. Specifically, we artificially remove 10 near-offset traces from datasets that originally contain these traces, train the model on the remaining incomplete data, and measure reconstruction accuracy against the known ground truth. This controlled validation strategy enables rigorous quantitative comparison with baseline methods, while providing detailed insights into reconstruction fidelity, amplitude preservation, phase consistency, and spectral characteristics. We compare our method against the parabolic Radon transform (PRT), a widely adopted benchmark for seismic trace interpolation, to demonstrate the advantages of our approach. 

Crucially, although ground-truth data are used for post-hoc evaluation, the training process itself operates in a self-supervised manner without access to complete near-offset references, thereby reflecting the realistic constraints of towed-streamer acquisition. Following these controlled validation experiments, we demonstrate operational deployment on actual near-offset gaps in the next section, where ground-truth validation is inherently impossible.

\subsection{Synthetic data}

The synthetic dataset is generated by forward modeling on the open-source 2D SEAM (SEG Advanced Modeling) velocity model using an acoustic finite-difference wave-equation simulation. We employ a Ricker wavelet with a dominant frequency of 10~Hz as the source signature. To mimic realistic towed-streamer geometry, we adopt a single-sided receiver configuration. A total of 882 shots are simulated with a shot interval of 10~m. Each shot gather contains 320 receivers with a spacing of 10~m, and the data are sampled at 1200 time steps with a temporal sampling interval of 4~ms, yielding a maximum recording time of approximately 4.8~s. The complete 882 shot gathers constitute our training dataset. To simulate the near-offset gap inherent in towed-streamer acquisition, we remove the first 10 near-offset traces from all shot gathers, which constitutes a gap of 100 meters. 

\begin{figure}[htbp]
\centering
\includegraphics[width=1\textwidth]{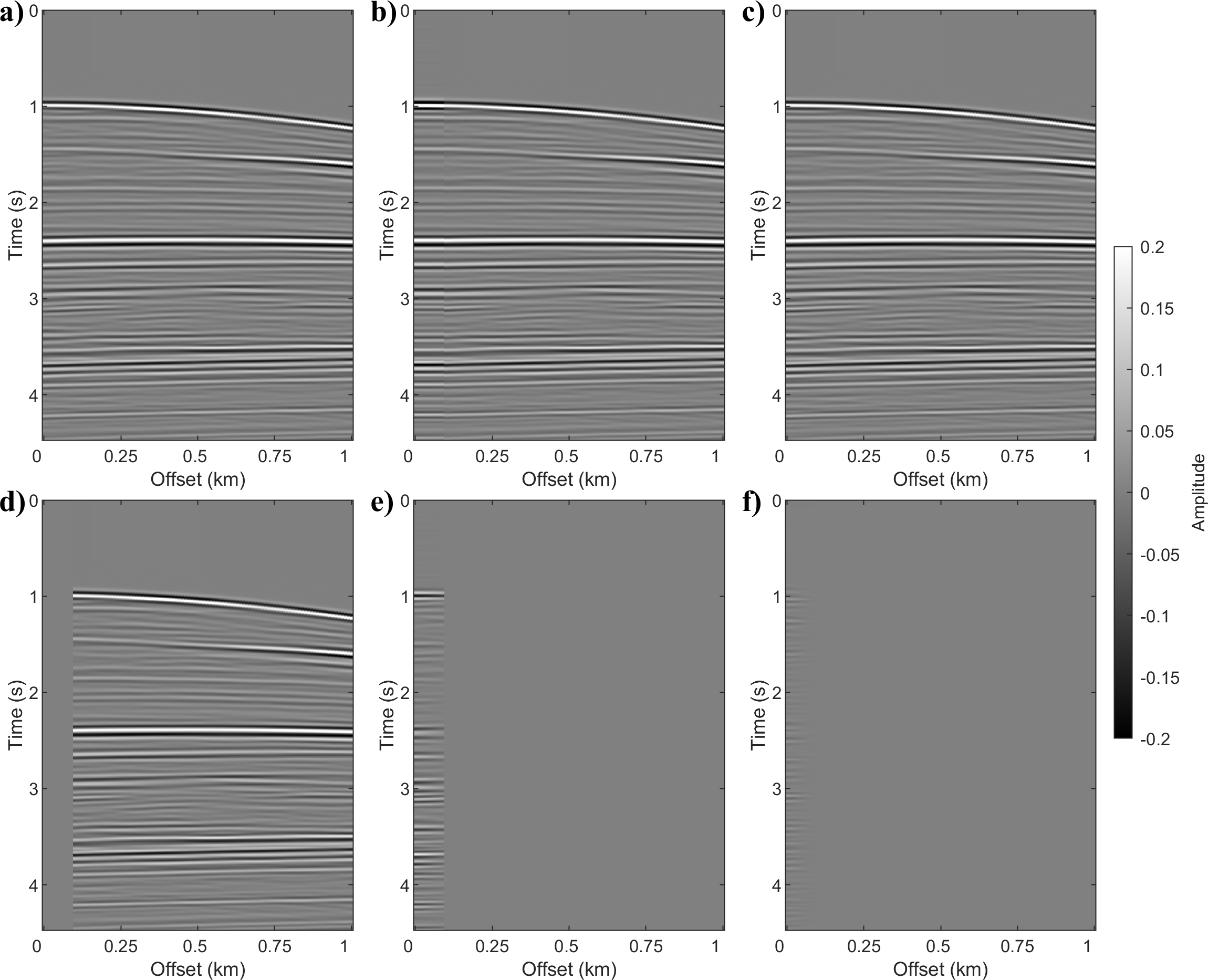}
\caption{Comparison of near-offset reconstruction results on synthetic data. The 101 traces closest to zero offset (0-1~km) are displayed. (a) Complete ground-truth shot gather. Reconstruction data by (b) the PRT and (c) our method. (d) Incomplete observed data with the first 10 near-offset traces removed. Reconstruction error (difference from ground truth) of (e) the PRT and (f) our method.}
\label{fig1}
\end{figure}

For training, the lateral patch width $W$ is set to 32 traces (see Equation \ref{eq7}), and patches are randomly extracted from the available far-offset region of each shot gather. The model is trained with a batch size of 16 and a fixed learning rate of $1 \times 10^{-4}$ using an AdamW optimizer. Training proceeds for 20000 iterations on a single NVIDIA RTX 8000 GPU, taking approximately 7 hours. The diffusion process employs $T=1000$ time steps during training, and DDIM sampling with $S=10$ steps is used during inference for computational efficiency.

Upon completion of training, we evaluate the reconstruction performance on a single shot gather selected from the dataset. Figure~\ref{fig1} compares the reconstruction results of PRT and our method, focusing on the 101 traces closest to zero offset (spanning approximately 0-1~km in offset) to highlight the near-offset reconstruction quality. Panel (a) displays the complete ground-truth data for these 101 traces, serving as the reference. Panel (d) shows the incomplete observed data with the first 10 near-offset traces removed, which is also the input for both methods. Panels (b) and (c) present the reconstructed shot gathers obtained by PRT and our method, respectively. The corresponding reconstruction errors, computed as the difference between the reconstructed data and ground-truth, are shown in panels (e) and (f). Visual inspection reveals that our method achieves significantly higher interpolation accuracy compared to PRT, effectively recovering the missing near-offset traces with improved kinematic and amplitude consistency. Notably, the interpolated traces from PRT exhibit noticeable phase misalignments and abrupt discontinuities at the boundary between the reconstructed and recorded traces, whereas our method produces a smooth and seamless transition with well-aligned seismic events across the entire offset range.

\begin{figure}[htbp]
\centering
\includegraphics[width=1\textwidth]{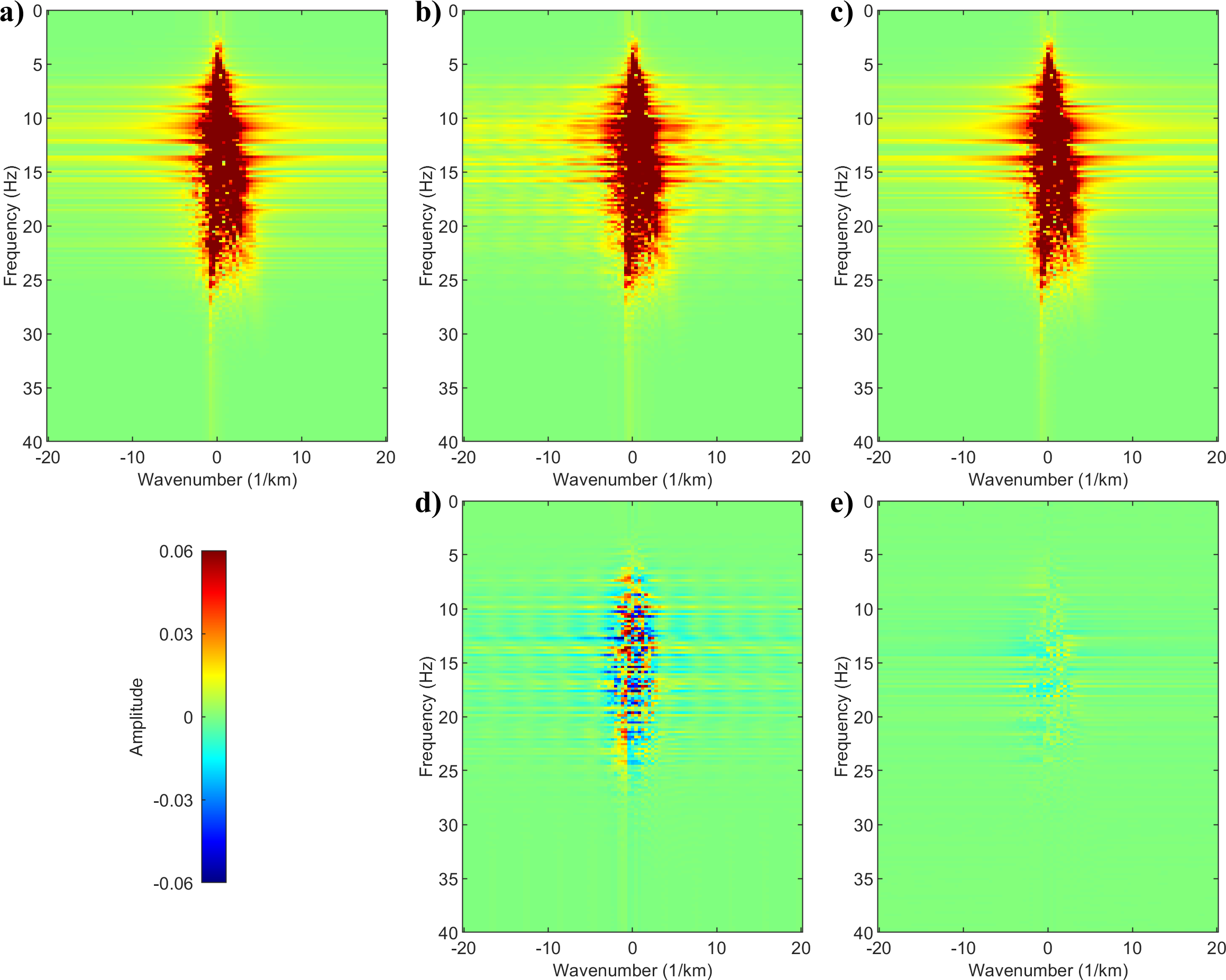}
\caption{Frequency-wavenumber (F-K) spectra comparison for the near-offset reconstruction on synthetic data. (a) F-K spectra of the complete ground-truth data. F-K spectra of the reconstruction data corresponding to (b) the PRT and (c) our method. Spectra error (difference from ground truth) of (d) the PRT and (e) our method. }
\label{fig2}
\end{figure}

To further assess the spectrum characteristics of the reconstructed data, we compute the frequency-wavenumber (F-K) spectrum and present the results in Figure~\ref{fig2}. Panel (a) displays the F-K spectra of the complete ground-truth data. Panels (b) and (c) show the F-K spectra of the reconstructed data obtained by PRT and our method, respectively, while panels (d) and (e) present the corresponding Spectra errors (differences from the ground-truth spectrum). Our method produces an F-K spectra that closely matches the ground truth, with minimal residual energy in the error panel, indicating excellent preservation of the spectrum content and spatial coherence of the seismic wavefield. In contrast, PRT introduces noticeable spectrum artifacts and fails to accurately recover the low-wavenumber components associated with near-offset traces, as evidenced by the larger residual energy in panel (d).

\begin{figure}[htbp]
\centering
\includegraphics[width=1\textwidth]{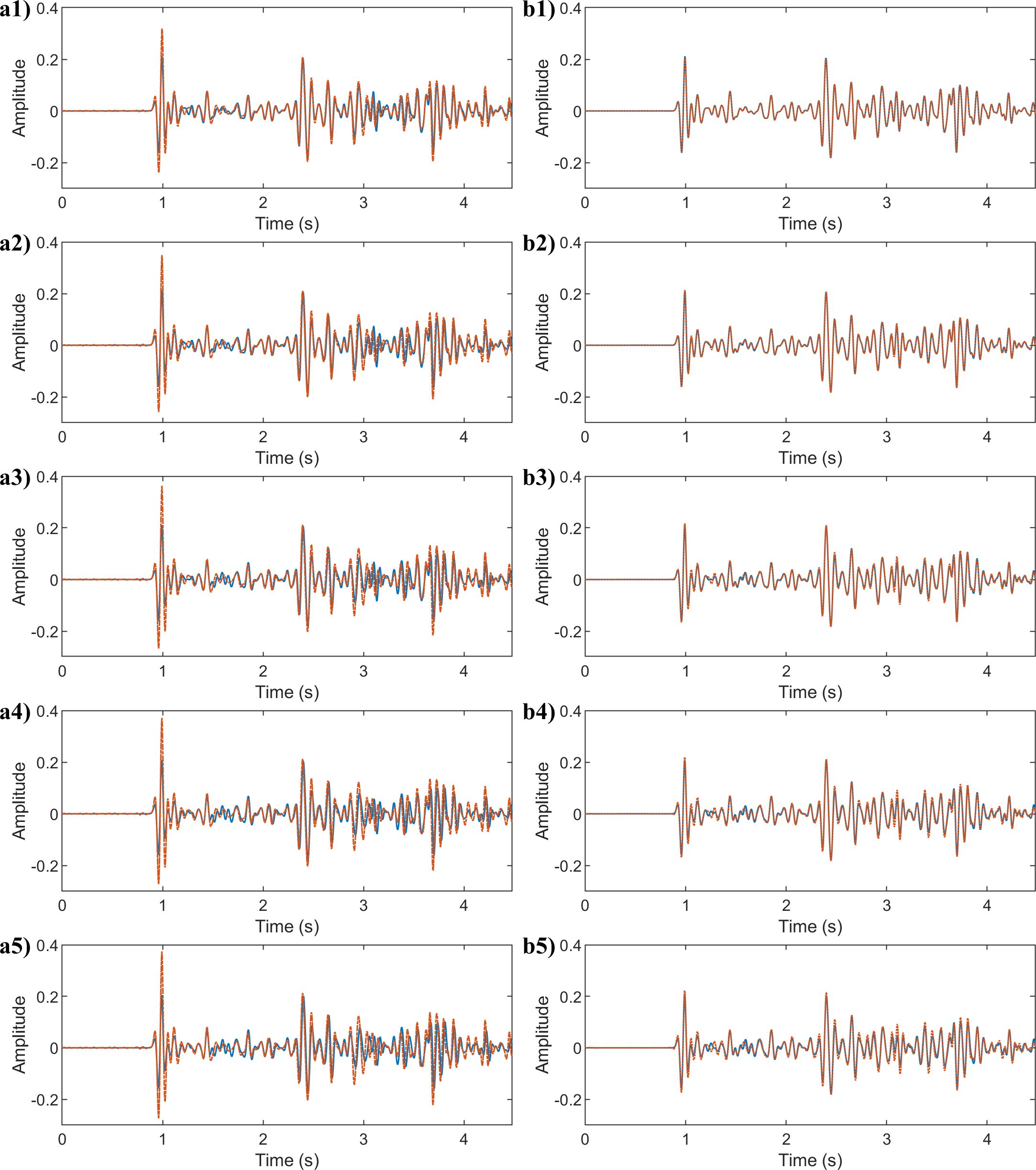}
\caption{Waveform comparison of reconstructed near-offset traces on synthetic data. Blue solid lines represent ground-truth waveforms, and orange dashed lines represent reconstructed waveforms. (a1-a5) Comparison of the 10th, 7th, 5th, 3rd, and 1st reconstructed traces (from far to near offset) obtained by PRT with ground truth. (b1-b5) Corresponding comparisons for our method. }
\label{fig3}
\end{figure}

For a more detailed waveform-level comparison, we extract five individual traces from the reconstructed near-offset section and compare them against the corresponding ground-truth traces in Figure~\ref{fig3}. Panels (a1)-(a5) display the 10th, 7th, 5th, 3rd, and 1st reconstructed traces (from far to near offset) obtained by PRT alongside the ground-truth waveforms, while panels (b1)-(b5) present the corresponding comparisons for our method. In each panel, the blue solid line represents the ground-truth waveform, and the orange dashed line represents the reconstructed waveform. The PRT exhibits substantial amplitude discrepancies across all traces, even at the 10th trace, which is closest to the available far-offset data. As the offset decreases (approaching zero offset), phase misalignments become increasingly pronounced, e.g., the 3rd and 1st traces, where significant errors are evident. Such phase errors can severely degrade subsequent processing steps, including velocity analysis and full-waveform inversion. In contrast, our method achieves excellent waveform matching across all traces. For the 10th to 5th traces, the reconstructed waveforms nearly perfectly align with the ground truth in both amplitude and phase. Even for the nearest-offset traces (3rd and 1st), where extrapolation uncertainty is highest, our method maintains remarkable fidelity, with only minor amplitude deviations and negligible phase errors on a small subset of events.

\begin{figure}[htbp]
\centering
\includegraphics[width=1\textwidth]{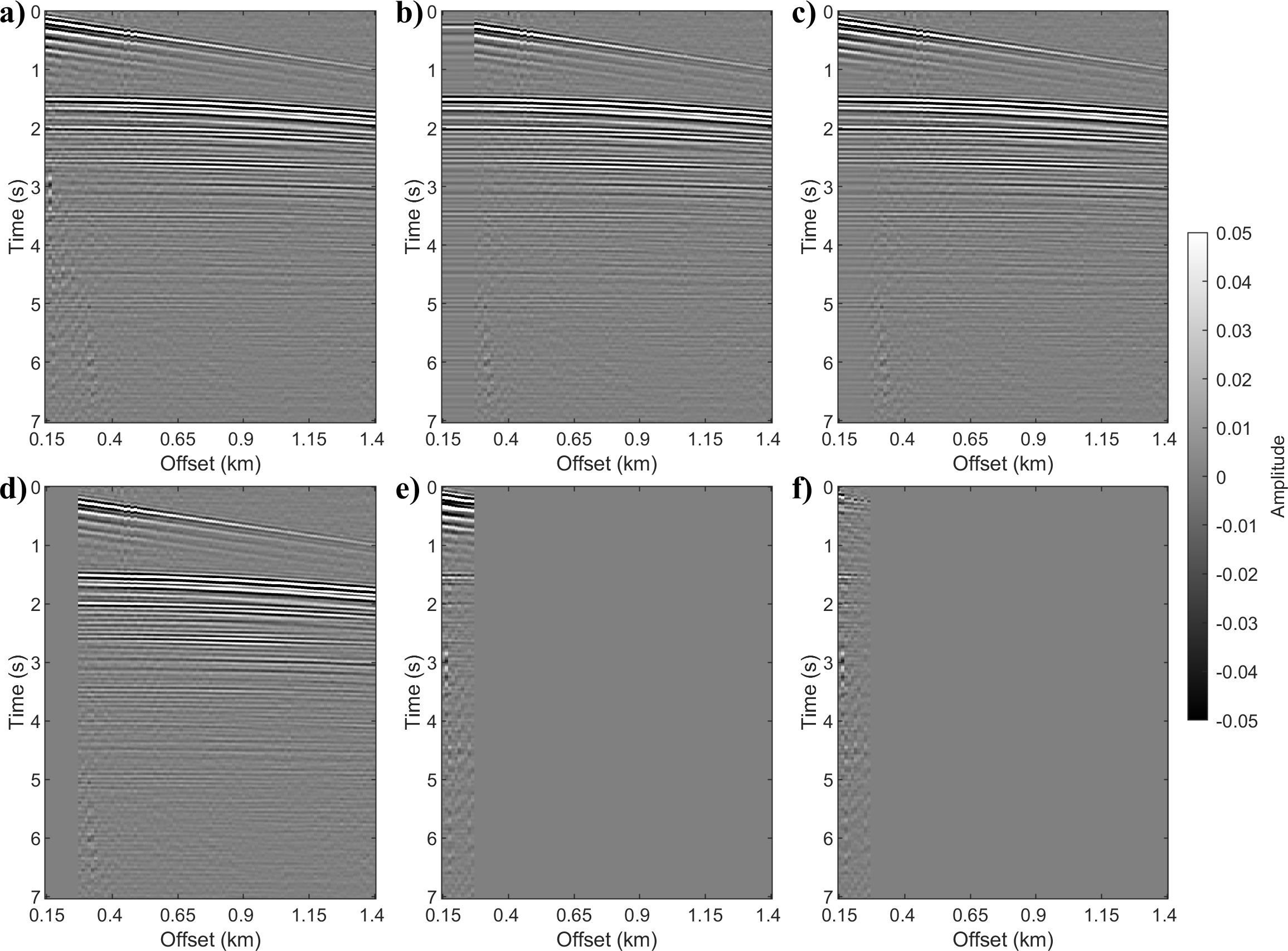}
\caption{Near-offset reconstruction results on the field data from northwest Australia for validation purposes. The 101 traces closest to zero offset are displayed. (a) Complete ground-truth (recorded traces starting from $\sim$150~m offset). Reconstruction data of 10 removed traces for validation purposes by (b) the PRT and (c) our method. (d) Incomplete data (10 traces artificially removed from recorded portion for validation). Reconstruction error (difference from ground truth) of (e) the PRT and (f) our method.}
\label{fig4}
\end{figure}

\begin{figure}[htbp]
\centering
\includegraphics[width=1\textwidth]{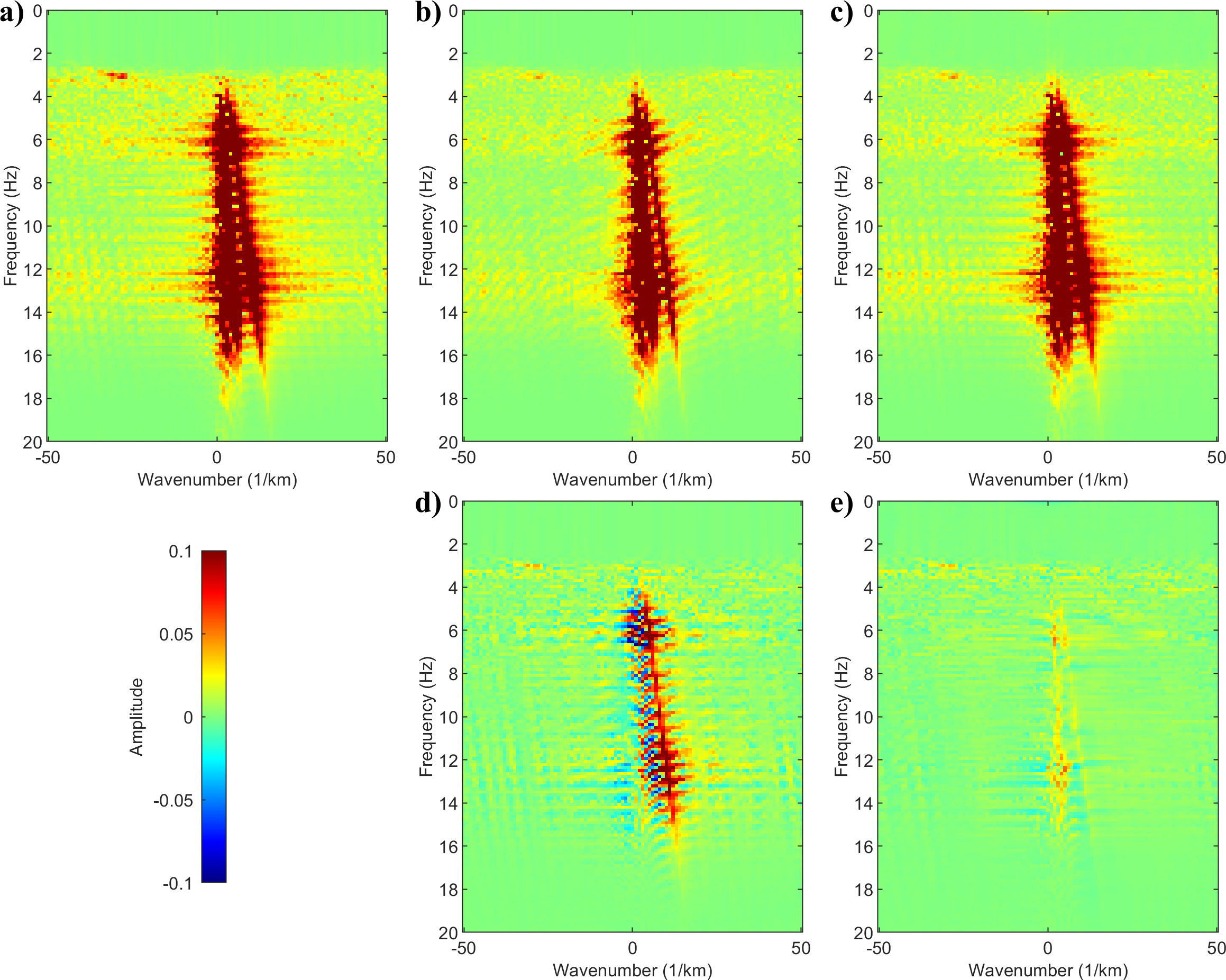}
\caption{F-K spectra comparison for field data I (northwest Australia) validation experiment. (a) F-K spectra of the ground-truth data (from recorded traces). F-K spectra of the reconstruction data corresponding to (b) the PRT and (c) our method. Spectra error (difference from ground truth) of (d) the PRT and (e) our method. }
\label{fig5}
\end{figure}

\begin{figure}[htbp]
\centering
\includegraphics[width=1\textwidth]{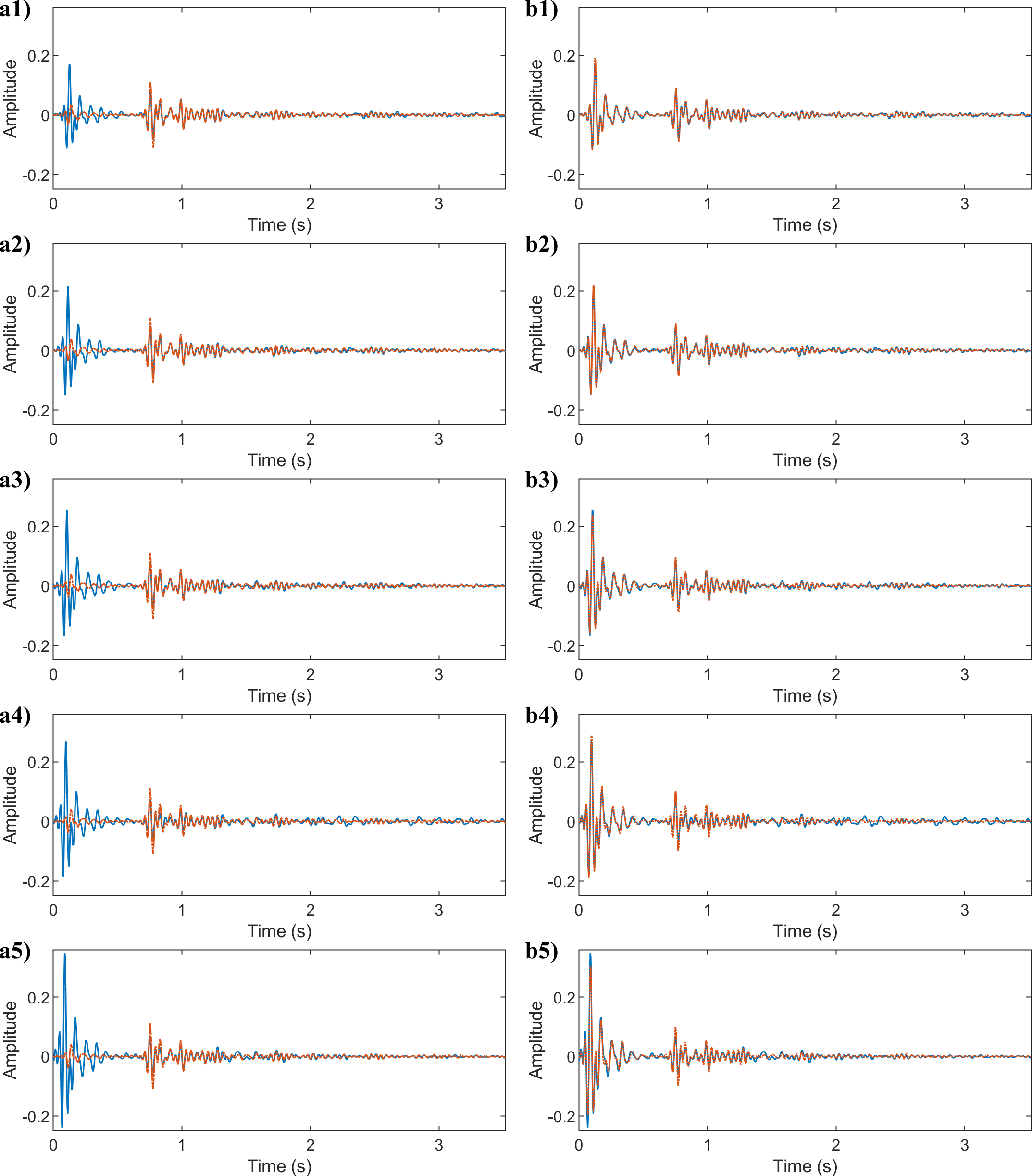}
\caption{Waveform comparison for field data I (northwest Australia) validation experiment. Blue solid lines: ground truth (recorded traces); Orange dashed lines: reconstruction. (a1-a5) PRT results for the 10th, 7th, 5th, 3rd, and 1st artificially removed traces (from far to near offset). (b1-b5) The proposed method results.}
\label{fig6}
\end{figure}

\subsection{Field data I: Northwest Australia survey}

The first field dataset was acquired offshore northwest Australia using a variable-depth towed-streamer system. The survey comprises 1824 shots with a shot interval of approximately 18.75~m. Each shot gather contains 648 receivers with a spacing of 12.5~m and 7040 time samples at a 1~ms sampling interval, yielding a maximum recording time of approximately 7~s. It is important to note that this dataset, like most towed-streamer acquisitions, already exhibits an inherent near-offset gap of approximately 150~m (corresponding to 12 traces at 12.5~m spacing) due to the physical source-receiver separation. However, for validation purposes in this controlled experiment, we utilize the recorded traces starting from 150~m offset as if they were complete data. We then artificially remove an additional 10 traces from this recorded portion (i.e., traces at 150-275~m offset) to simulate a missing near-offset zone, which enables quantitative performance evaluation against pseudo ground-truth references. The reconstruction of the actual 0-150~m gap will be demonstrated in the next section.

We use the first 1000 shots as the training dataset. Here, an important consideration is that during training, when randomly extracting patches from each shot gather at every iteration (see Equation~\ref{eq7}), we restrict the training patch extraction to the first 200 traces (measured from the start of the recorded data at 150~m offset) rather than utilizing the entire offset aperture. This restriction is motivated by the observation that far-offset traces exhibit significantly larger amplitude variations compared to near-offset traces due to geometric spreading, which can dominate the training loss and hinder the model's ability to learn subtle amplitude trends relevant to near-offset reconstruction.

Rather than training from random initialization, we employ transfer learning by initializing the network with the model trained on the synthetic dataset. This warm-start strategy substantially accelerates convergence, reducing the required training iterations to 8000 (compared to 20000 for synthetic data) and the total training time to approximately 2 hours and 21 minutes on a single NVIDIA RTX 8000 GPU. All other training hyperparameters (batch size, learning rate, optimizer, diffusion steps) remain identical to those used for synthetic data.

Following the same evaluation as the synthetic experiment, we assess reconstruction performance on a selected test shot gather. Figure~\ref{fig4} presents the reconstruction comparison for the 101 traces closest to zero offset (spanning approximately 0.15-1.4~km offset). Panel (a) shows the complete ground-truth shot gather (recorded traces used as reference), panel (d) displays the incomplete observed data with 10 artificially removed near-offset traces, and panels (b) and (c) show the reconstructions obtained by PRT and our method, respectively. The reconstruction errors are presented in panels (e) and (f). Consistent with the synthetic results, our approach achieves markedly superior reconstruction fidelity compared to PRT. Examination of the error panels reveals that the PRT (panel e) exhibits significant signal leakage, with residual energy from both direct arrivals and primary reflections clearly visible. In contrast, our method (panel f) achieves nearly artifact-free reconstruction with minimal signal leakage. Notably, both methods demonstrate a degree of noise suppression capability, effectively attenuating acquisition noise while recovering the near-offset traces.

The F-K spectra comparison in Figure~\ref{fig5} further corroborates these observations. Our method (panel c) closely reproduces the spectrum characteristics of the ground truth (panel a), with minimal residual energy in the error panel (e). In contrast, PRT (panel b) exhibits spectrum distortions, particularly in the low-wavenumber region corresponding to near-offset events, as evidenced by the larger residual in panel (d). These spectrum artifacts can propagate into subsequent processing steps and degrade imaging quality.

Detailed waveform comparisons for the 10th, 7th, 5th, 3rd, and 1st reconstructed traces are presented in Figure~\ref{fig6}. The PRT (panels a1-a5) shows consistent amplitude over- or under-estimation across all traces, with increasing phase errors toward smaller offsets. The proposed method (panels b1-b5) achieves excellent waveform fidelity for traces 10-5, with only minor amplitude deviations appearing at the 3rd and 1st traces where extrapolation becomes more challenging. Importantly, the phase alignment remains highly accurate even at the nearest offsets, ensuring that the reconstructed near-offset traces can reliably support subsequent velocity inversion workflows that do not rely too much on the amplitude.

\begin{figure}[htbp]
\centering
\includegraphics[width=1\textwidth]{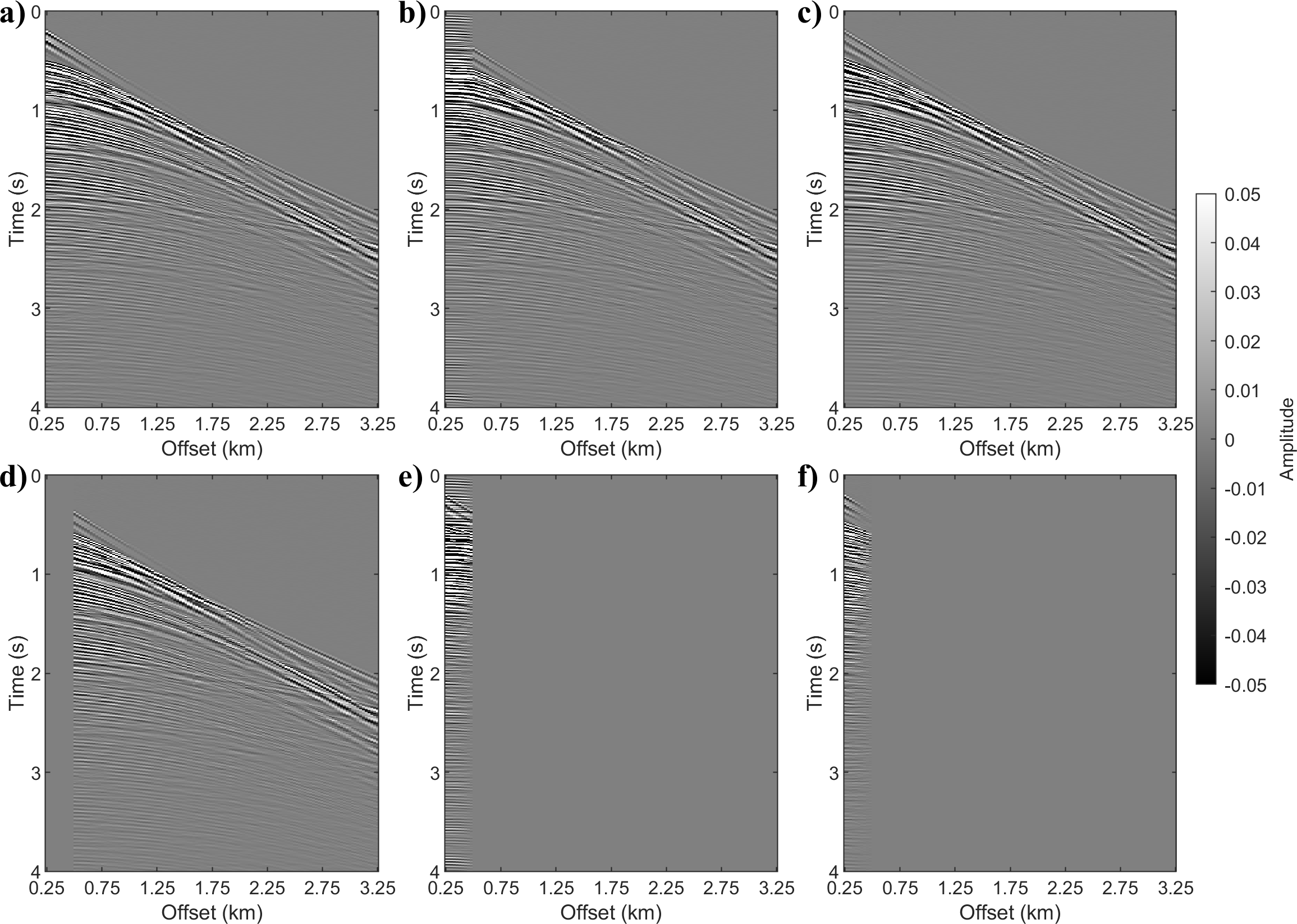}
\caption{Near-offset reconstruction results on the Mobil AVO viking graben line 12 dataset for validation purposes. The 101 traces closest to zero offset are displayed. (a) Complete ground-truth (recorded traces starting from $\sim$250~m offset). Reconstruction data of 10 removed traces for validation purposes by (b) the PRT and (c) our method. (d) Incomplete data (10 traces artificially removed from recorded portion for validation). Reconstruction error (difference from ground truth) of (e) the PRT and (f) our method.}
\label{fig7}
\end{figure}

\begin{figure}[htbp]
\centering
\includegraphics[width=1\textwidth]{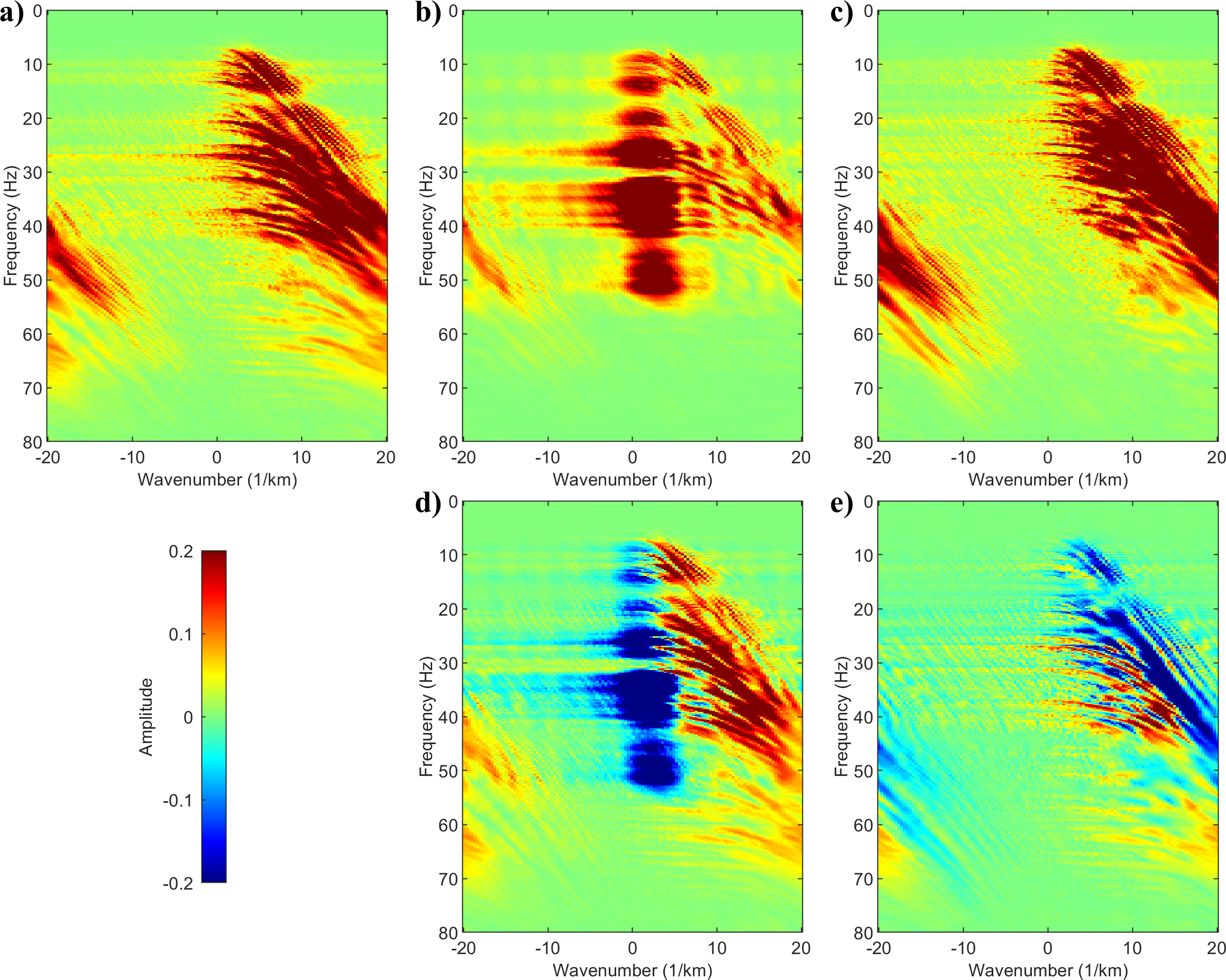}
\caption{F-K spectra comparison for field data II (Mobil AVO viking graben line 12) validation experiment. (a) F-K spectra of the ground-truth data (from recorded traces). F-K spectra of the reconstruction data corresponding to (b) the PRT and (c) our method. Spectra error (difference from ground truth) of (d) the PRT and (e) our method. }
\label{fig8}
\end{figure}

\begin{figure}[htbp]
\centering
\includegraphics[width=1\textwidth]{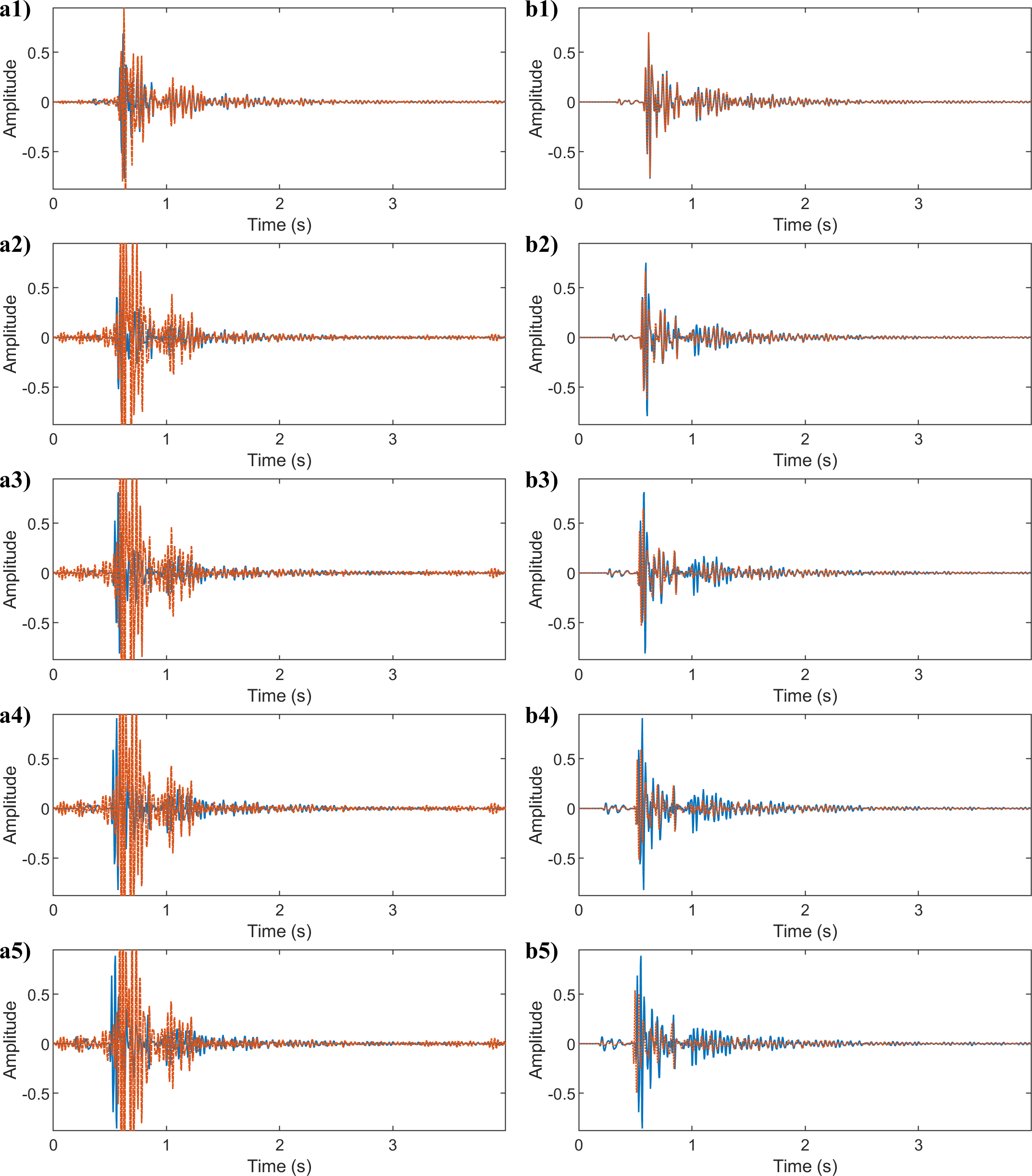}
\caption{Waveform comparison for field data II (Mobil AVO viking graben line 12) validation experiment. Blue solid lines: ground truth (recorded traces); Orange dashed lines: reconstruction. (a1-a5) PRT results for the 10th, 7th, 5th, 3rd, and 1st artificially removed traces (from far to near offset). (b1-b5) The proposed method results.}
\label{fig9}
\end{figure}

\subsection{Field data II: Mobil AVO Viking Graben Line 12}

The second field dataset is the publicly available SEG Mobil AVO viking graben line 12, acquired with towed-streamer geometry. This marine survey consists of 1001 shots, each containing 120 receivers. Both the shot interval and receiver spacing are 25~m. The recording length is approximately 6~s with a temporal sampling interval of 4~ms. Similar to field dataset I, this acquisition inherently lacks a near-offset zone of approximately 250~m (10 traces at 25~m spacing) due to source-receiver separation. For validation purposes, we follow the same strategy as described in previous field experiment: we treat the recorded traces starting from 250~m offset as reference data and artificially remove an additional 10 traces from this recorded portion to enable quantitative evaluation against ground truth. The actual 0-250~m gap will be reconstructed in the next section.

We use the complete 1001-shot dataset for training, truncating each shot gather to the first 4~s to focus on the primary reflection zone. This dataset presents a significantly more challenging reconstruction scenario compared to the synthetic and first field examples. Visual inspection of the complete data (Figure~\ref{fig7}a) reveals pronounced spatial aliasing due to the relatively coarse receiver spacing (25~m), resulting in poor lateral continuity of seismic events, particularly for steeply dipping arrivals and high-frequency components. Training configuration follows the same hyperparameter settings as the synthetic experiment, with transfer learning initialization from the synthetic-trained model. The network is trained for 20000 iterations, requiring approximately 6 hours on a single NVIDIA RTX 8000 GPU. 

Figure~\ref{fig7} presents the reconstruction comparison, displaying 120 traces approximately spanning 0.25-3.25~km in offset. Compared to the PRT baseline, our method demonstrates superior reconstruction fidelity with notably reduced signal leakage visible in the error panels (e) and (f). The PRT reconstruction (panel b) exhibits a critical failure mode: the reconstructed near-offset events appear nearly horizontal, exhibiting severe kinematic inconsistency with the recorded far-offset data. This manifests as extremely poor lateral continuity at the reconstruction boundary and is particularly pronounced for high-amplitude shallow events (time $<1$~s), where the PRT introduces strong artifacts that show little resemblance to the true wavefield structure. These artifacts arise from the method's inability to properly handle the spatially aliased data, leading to wrong event mapping in the Radon domain. In contrast, our method (panel c) achieves significantly better lateral continuity with the recorded data, maintaining coherent event alignment across the reconstruction boundary. However, both methods exhibit more substantial signal leakage than observed in the previous two experiments, reflecting the inherent difficulty of reconstructing aliased near-offset data from coarsely sampled far-offset observations.

The F-K spectra analysis in Figure~\ref{fig8} further quantifies these observations. The PRT spectrum (panel b) deviates significantly from the ground-truth spectrum (panel a), with the error panel (d) revealing severe spectrum distortion across a wide range of frequencies and wavenumbers. This indicates that the PRT fails to recover the fundamental spectrum characteristics of the near-offset wavefield in the presence of strong aliasing. Our method (panel c) achieves substantially better spectrum reconstruction, producing a spectrum that more closely resembles the ground truth, despite with visible residual errors (panel e) that reflect the challenging nature of this dataset. The spectrum leakage in our reconstruction is acceptable and considerably smaller than that of the PRT, confirming improved preservation of spatial coherence despite the aliasing artifacts.

The waveform-level comparison in Figure~\ref{fig9} provides the most solid evidence of the PRT's limitations on this challenging dataset. For strong-amplitude reflection events, the PRT reconstruction (panels a1-a5) exhibits obvious errors even at the 10th trace, which is immediately adjacent to the recorded data. Massive amplitude discrepancies and severe phase misalignments are evident across all traces, rendering the reconstructed traces unsuitable for subsequent processing workflows. In stark contrast, our method (panels b1-b5) achieves excellent phase alignment for traces 10-5, with only minor amplitude deviations that preserve the overall waveform character. As the reconstruction progresses toward zero offset (traces 3 and 1), extrapolation errors increase, and some strong reflection events begin to exhibit phase shifts. Nevertheless, the overall reconstruction quality remains far superior to the PRT, particularly in maintaining kinematic consistency. These results demonstrate that while the proposed framework faces challenges when confronted with spatially aliased field data, it maintains substantially better robustness and reconstruction fidelity compared to conventional transform-domain methods.

\subsection{Quantitative performance comparison}

To provide a comprehensive quantitative assessment of reconstruction quality, we evaluate both the PRT and the proposed method using three widely adopted metrics: mean absolute error (MAE), structural similarity index measure (SSIM), and signal-to-noise ratio (SNR). These metrics collectively assess different aspects of reconstruction fidelity.

The MAE measures the average absolute difference between the reconstructed and ground-truth near-offset data:
\begin{equation}
\text{MAE} = \frac{1}{N}\sum_{i=1}^{N}|\hat{x}_i - x_i|,
\end{equation}
where $\hat{x}_i$ and $x_i$ denote the reconstructed and ground-truth samples, respectively, and $N$ is the total number of samples in the near-offset region. Lower MAE values indicate better amplitude accuracy.

The SSIM evaluates perceptual similarity between reconstructed and ground-truth data by jointly considering luminance, contrast, and structural information \citep{wang2004image}:
\begin{equation}
\text{SSIM}(\hat{\mathbf{x}}, \mathbf{x}) = \frac{(2\mu_{\hat{\mathbf{x}}}\mu_{\mathbf{x}} + C_1)(2\sigma_{\hat{\mathbf{x}}\mathbf{x}} + C_2)}{(\mu_{\hat{\mathbf{x}}}^2 + \mu_{\mathbf{x}}^2 + C_1)(\sigma_{\hat{\mathbf{x}}}^2 + \sigma_{\mathbf{x}}^2 + C_2)},
\end{equation}
where $\mu$ and $\sigma$ denote mean and standard deviation, $\sigma_{\hat{\mathbf{x}}\mathbf{x}}$ represents covariance, and $C_1$, $C_2$ are small constants for numerical stability. SSIM ranges from $-1$ to $1$, with higher values indicating better structural preservation.

The SNR quantifies the ratio between signal power and reconstruction error power:
\begin{equation}
\text{SNR} = 10\log_{10}\left(\frac{\sum_{i=1}^{N}x_i^2}{\sum_{i=1}^{N}(\hat{x}_i - x_i)^2}\right) \text{ (dB)}.
\end{equation}
Higher SNR values correspond to lower reconstruction errors relative to the signal strength.

Table~\ref{tab:metric_comparison} summarizes the quantitative comparison across the three test datasets. For the synthetic dataset, our method achieves substantial improvements over the PRT: MAE is reduced by 81.5\% (from 0.0178 to 0.0033), SSIM increases from 0.6727 to 0.9597, and SNR improves dramatically from 2.52~dB to 13.62~dB. For field dataset I (northwest Australia), our method maintains significant performance advantages: MAE decreases by 54.5\% (from 0.0088 to 0.0040), SSIM increases from 0.8355 to 0.9455, and SNR improves from 0.20~dB to 10.92~dB. For field dataset II (Mobil AVO viking graben line 12), which presents the most challenging reconstruction scenario due to pronounced spatial aliasing, our method achieves lower MAE (0.0214 vs. 0.0705) and higher SSIM (0.8063 vs. 0.5419) compared to the PRT. However, both methods yield negative SNR values (-1.6655~dB for ours vs. -9.2979~dB for PRT), indicating that reconstruction errors exceed the signal power in the near-offset region. This reflects the fundamental difficulty of recovering aliased near-offset information from coarsely sampled far-offset observations. Nevertheless, our method's SNR represents an 8~dB improvement over the PRT, and the substantially lower MAE confirms more accurate reconstruction even under severe aliasing conditions.

Overall, the quantitative metrics consistently demonstrate that the proposed method outperforms the conventional baseline across diverse datasets, with performance gains ranging from moderate improvements on challenging aliased data to significant enhancements on well-sampled examples.

\begin{table}[t]
\centering
\caption{Comparison of MAE, SSIM, and SNR between PRT and our method across three datasets, with bold text indicating best accuracy.}
\label{tab:metric_comparison}
\begin{tabular}{lccccc}
\hline
\textbf{Dataset} & \textbf{Method} & \textbf{MAE} & \textbf{SSIM} & \textbf{SNR} \\
\hline
\multirow{2}{*}{Synthetic dataset} 
    & PRT & 0.0178 & 0.6727 & 2.5237\\
    & Our method & \textbf{0.0033} & \textbf{0.9597} & \textbf{13.6237} \\
\hline
\multirow{2}{*}{Field dataset I} 
    & PRT & 0.0088 & 0.8355 & 0.1962\\
    & Our method & \textbf{0.0040} & \textbf{0.9455} & \textbf{10.9246} \\
\hline
\multirow{2}{*}{Field dataset II} 
    & PRT & 0.0705 & 0.5419 & -9.2979\\
    & Our method & \textbf{0.0214} & \textbf{0.8063} & \textbf{-1.6655} \\
\hline
\end{tabular}
\end{table}

\begin{figure}[htbp]
\centering
\includegraphics[width=0.75\textwidth]{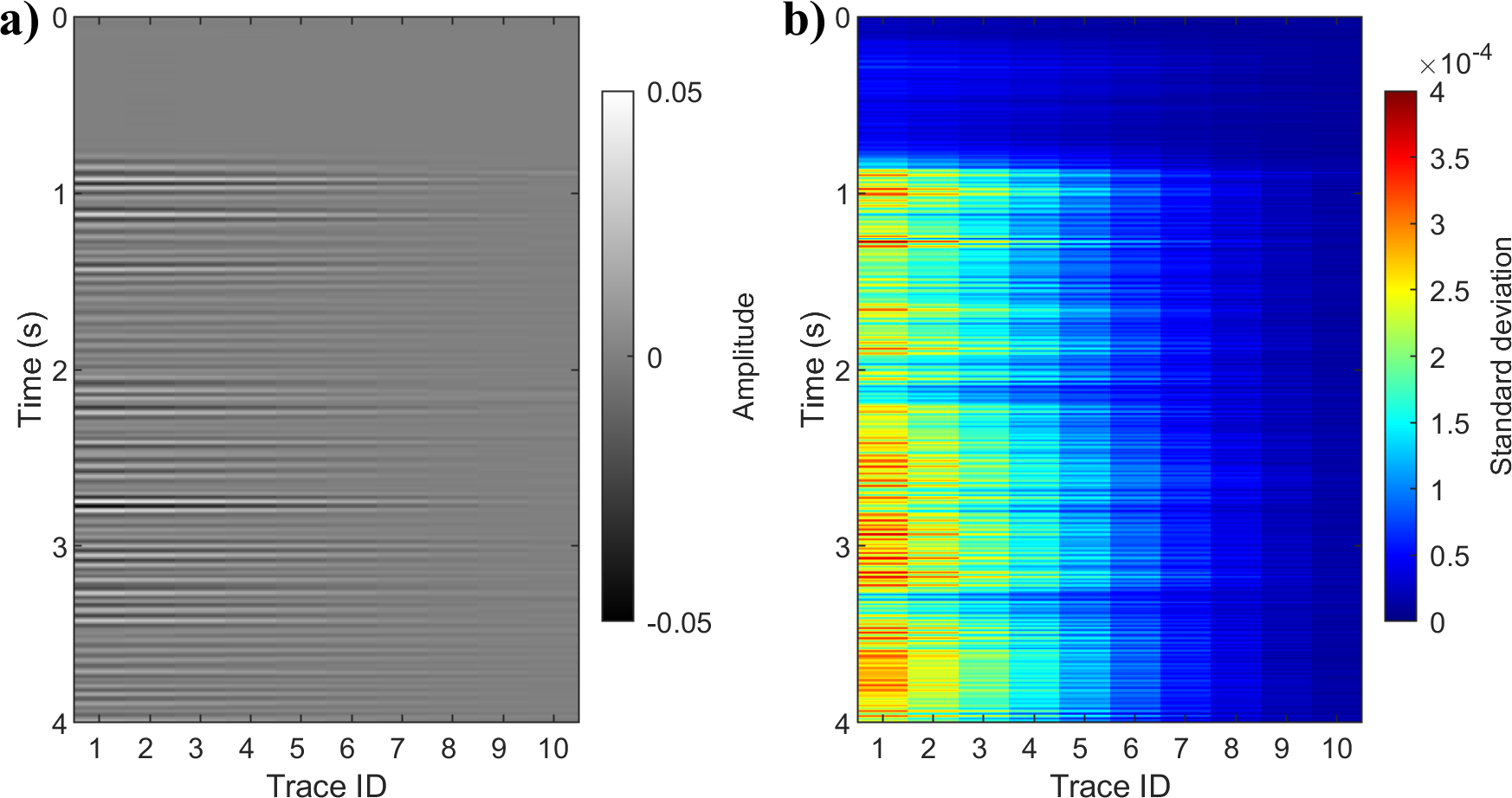}
\caption{Uncertainty quantification for the synthetic dataset. (a) Reconstruction error for the first 10 reconstructed near-offset traces (mean of 50 realizations minus ground truth). (b) Uncertainty map (standard deviation across 50 realizations). }
\label{fig10}
\end{figure}

\begin{figure}[htbp]
\centering
\includegraphics[width=0.75\textwidth]{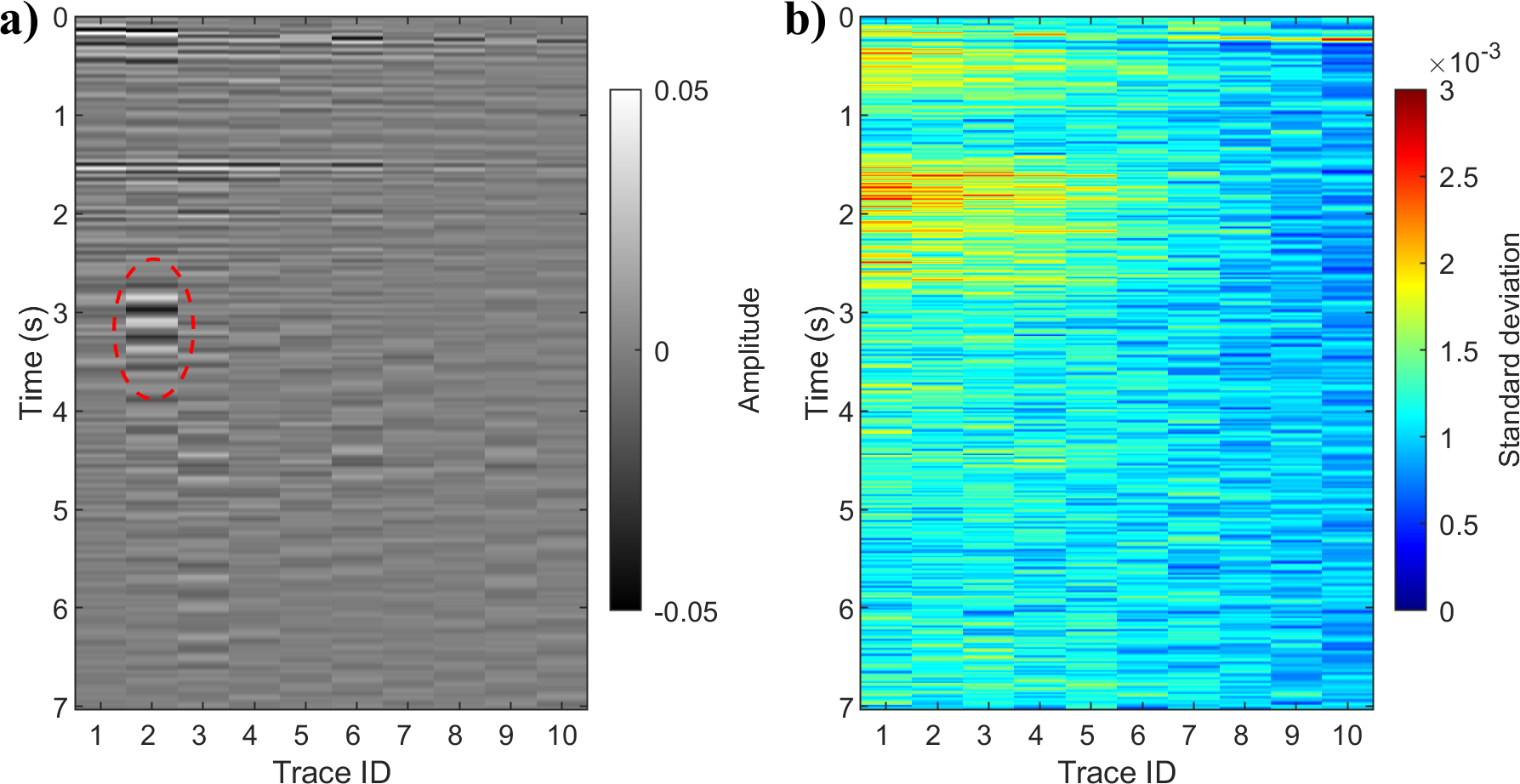}
\caption{Similar to Figure \ref{fig10}, but for field dataset I (northwest Australia).}
\label{fig11}
\end{figure}

\subsection{Uncertainty quantification analysis}

A particular advantage of the proposed diffusion-based framework over deterministic reconstruction methods is its natural capability for uncertainty quantification through ensemble sampling. To assess the reliability and spatial distribution of reconstruction uncertainty, we generate 50 independent realizations for each missing near-offset trace by performing DDIM sampling with different random noise initializations. The pointwise standard deviation across these realizations provides a pixel-wise uncertainty map that quantifies the model's confidence in the reconstruction. 

\begin{figure}[htbp]
\centering
\includegraphics[width=0.75\textwidth]{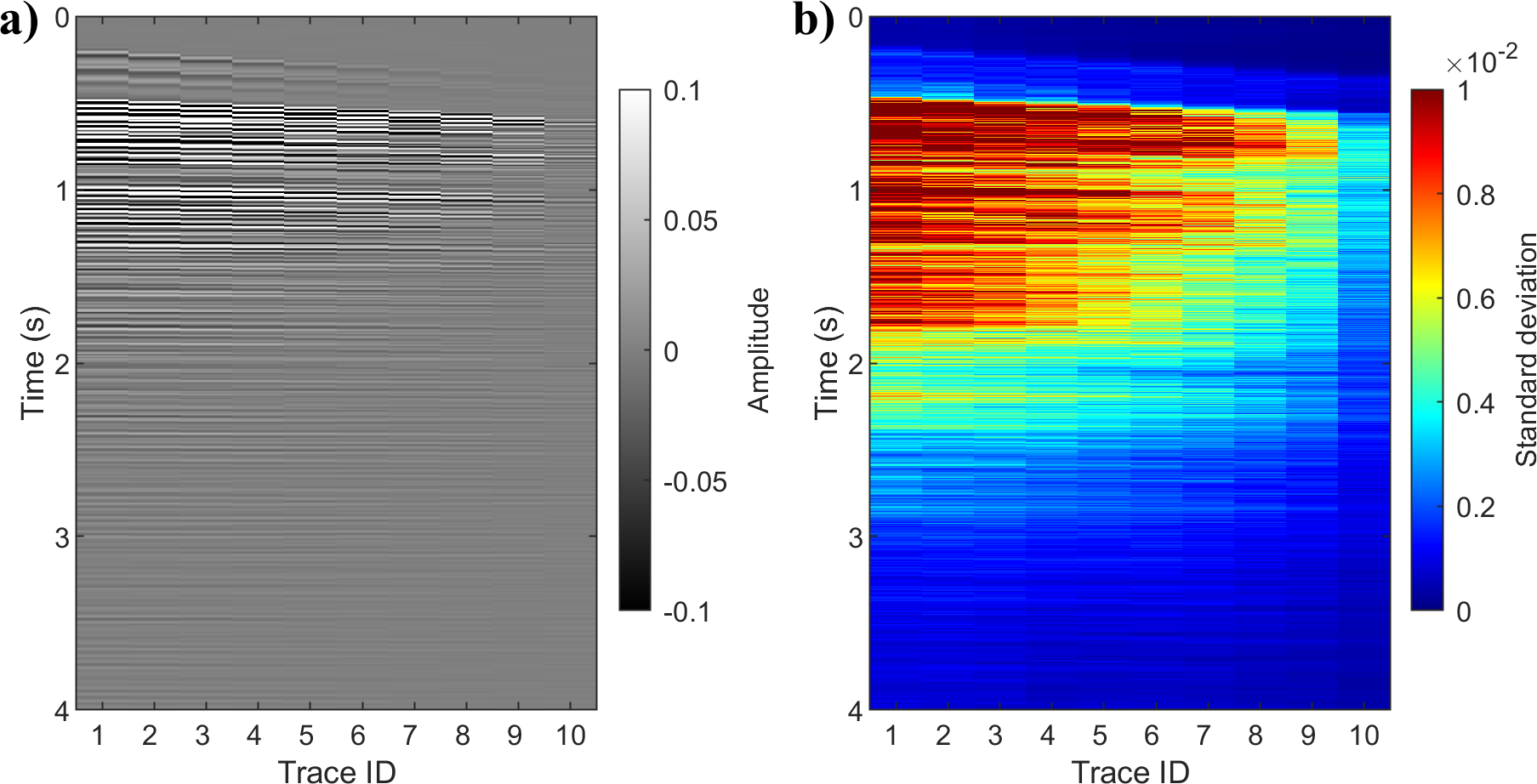}
\caption{Similar to Figure \ref{fig10}, but for field dataset II (Mobil AVO viking graben line 12).}
\label{fig12}
\end{figure}

We, here, focus our analysis on the first 10 reconstructed traces (closest to zero offset). Figures~\ref{fig10}, \ref{fig11}, and \ref{fig12} present the uncertainty quantification results for the synthetic dataset, field dataset I (northwest Australia), and field dataset II (Mobil AVO viking graben line 12), respectively. For each dataset, we display the reconstruction error (difference between the mean reconstruction and ground truth, panel a) alongside the corresponding uncertainty map (standard deviation across 50 realizations, panel b) for the 10 reconstructed near-offset traces. 

Several consistent and revealing patterns emerge across all three datasets. First, the spatial distribution of uncertainty exhibits strong correlation with reconstruction error: regions with large prediction errors (e.g., poorly reconstructed events) consistently display elevated uncertainty values, while accurately reconstructed portions of the wavefield exhibit low uncertainty. This correlation validates that the ensemble variance exactly reflects reconstruction confidence. Second, both reconstruction error and uncertainty increase generally as offset decreases from the 10th trace (farthest from zero offset) toward the 1st trace (nearest to zero offset). This trend confirms our earlier qualitative observations from the waveform comparisons (Figures~\ref{fig3}, \ref{fig6}, and \ref{fig9}): extrapolation becomes progressively more challenging and less constrained as the distance from the available recorded data increases. The 10th trace, which is immediately adjacent to the recorded far-offset region, exhibits minimal uncertainty, while the 1st trace, which is furthest from any conditioning information, shows the highest uncertainty.

However, a particularly informative exception is observed in the region marked by the red dashed ellipse in Figure~\ref{fig11}a. This region displays elevated reconstruction error but notably low uncertainty in Figure~\ref{fig11}b. Careful inspection of the original ground-truth data in Figure~\ref{fig4}a reveals that this zone contains significant acquisition noise. The elevated error in this region actually reflects successful noise suppression by the proposed method: the reconstruction consistently produces cleaner traces across all 50 realizations (hence low uncertainty), but these denoised predictions differ from the noisy ground truth (hence elevated error). This phenomenon provides critical validation of the uncertainty quantification's reliability, where it measures the model's prediction variability rather than simply tracking differences from ground truth.

The demonstrated relationship between reconstruction error and uncertainty, including the informative behavior in noise-dominated regions, highlights the critical practical value of uncertainty quantification in real-world seismic processing workflows. In realistic towed-streamer acquisition, complete near-offset ground-truth data are unavailable, making it impossible to compute reconstruction errors or directly assess interpolation accuracy. However, the uncertainty maps generated by our method provide an effective alternative means of evaluating reconstruction reliability without requiring ground truth. High-uncertainty regions can be flagged for manual inspection, excluded from subsequent processing steps, or subjected to alternative reconstruction strategies, while low-uncertainty regions can be confidently incorporated into surface-related multiple elimination and imaging workflows. Importantly, the demonstrated ability to distinguish between stochastic prediction variability (high uncertainty) and deterministic denoising behavior (low uncertainty despite error) further enhances the practical utility of these uncertainty estimates. This self-assessment capability distinguishes the proposed diffusion-based approach from conventional deterministic methods and significantly enhances its applicability.

\section{\textbf{Real-world application to actual near-offset gaps}}
In the preceding validation experiments, we evaluated the proposed framework by artificially removing near-offset traces from recorded field data, enabling quantitative performance assessment against ground-truth references. This controlled evaluation strategy provided clear evidence of the method's superior reconstruction fidelity compared to conventional baselines. However, the primary motivation for developing this framework is to address the near-offset gaps that are inherent in towed-streamer acquisitions due to physical source-receiver separation, where ground-truth validation is fundamentally impossible.

In this section, we demonstrate the operational deployment of the proposed method on the actual missing near-offset zones in the two field datasets. Unlike the validation experiments, where we artificially removed additional traces from the recorded portion, we now train the models directly on the complete observed data as acquired and, then, reconstruct the absent near-offset zones. Without ground truth for validation, uncertainty quantification becomes the primary means of assessing reconstruction reliability.

\subsection{Application to field data I}
Field dataset I (northwest Australia) exhibits an inherent near-offset gap of approximately 150~m, which amounts to 12 missing traces at 12.5~m receiver spacing. Therefore, our goal here is to reconstruct the actual 0-150~m near-offset gap (12 missing traces). We retrain model using the complete observed data starting from 150~m offset, following the same training configuration established in validation experiments. 

\begin{figure}[htbp]
\centering
\includegraphics[width=\textwidth]{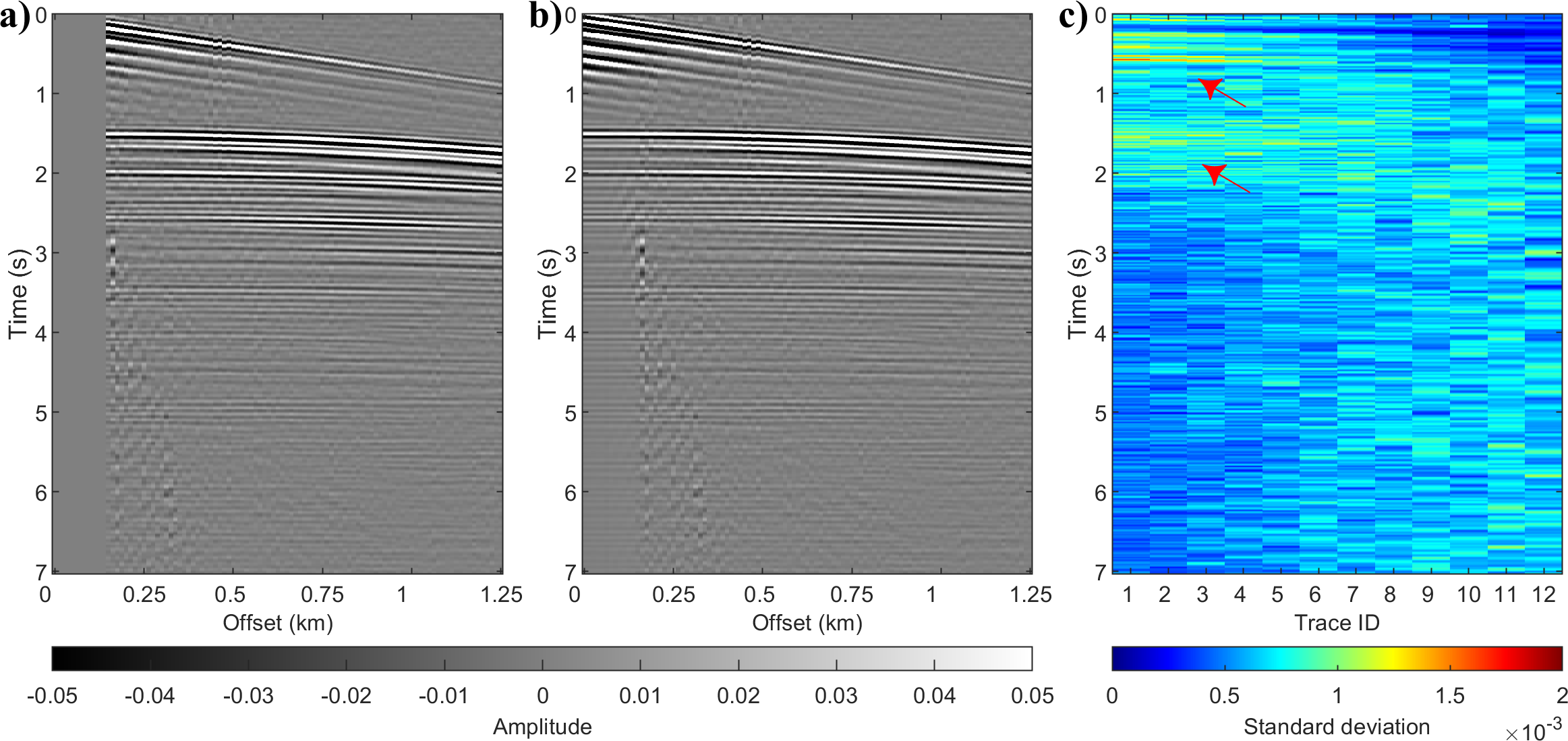}
\caption{Real near-offset gap reconstruction for field data I (northwest Australia). (a) Test shot gather displaying the complete offset range 0-1.25~km, where recorded data begin at $\sim$0.15~km (trace 13). (b) Reconstructed complete section including the 12 newly generated near-offset traces (0-150~m). (c) Uncertainty map (standard deviation across 50 realizations) for the 12 reconstructed traces.}
\label{fig13}
\end{figure}

Figure~\ref{fig13} presents the reconstruction results for the actual 0-150~m near-offset gap. Panel (a) shows a test shot gather displaying the complete offset range from 0 to 1.25~km, where the recorded data begin at approximately 0.15~km (trace 13). Panel (b) displays the reconstructed complete section including the 12 newly generated near-offset traces, and panel (c) presents the uncertainty map quantified by the standard deviation across 50 realizations. Visual inspection reveals several encouraging characteristics of the reconstruction. First, the reconstructed traces (0-150~m) exhibit excellent lateral continuity with the recorded traces ($>150$~m), with smooth event transitions across the reconstruction boundary and no visible kinematic discontinuities. The reconstructed section is remarkably clean, effectively avoiding noise contamination that might be present in the recorded far-offset data (consistent with the noise suppression behavior observed in the validation experiments). Second, the reconstructed direct arrival at zero offset emerges at approximately zero time, which is physically consistent with the expected source-receiver geometry at zero offset. This kinematic accuracy validates that the model has successfully learned and extrapolated the moveout behavior from the far-offset training data to the near-offset region. The uncertainty map (panel c) provides valuable insights into reconstruction confidence. Two localized zones exhibit elevated uncertainty, as indicated by the red arrows in panel (c): one near 1~s and another near 2~s, both concentrated in traces 1-4. In surface seismic recordings, shallow near-offset reflections often admit the largest curvature, and thus this might explain the higher uncertainty in the extrapolation. These regions correspond to shallow high-amplitude direct arrivals and reflection events. The elevated uncertainty correctly identifies these zones as requiring additional inspection, demonstrating the practical value of uncertainty quantification for quality control in the absence of ground-truth validation. Conversely, traces 8-12 and deeper events ($>3$~s) display lower uncertainty, suggesting higher reconstruction confidence in these regions.

\begin{figure}[htbp]
\centering
\includegraphics[width=0.75\textwidth]{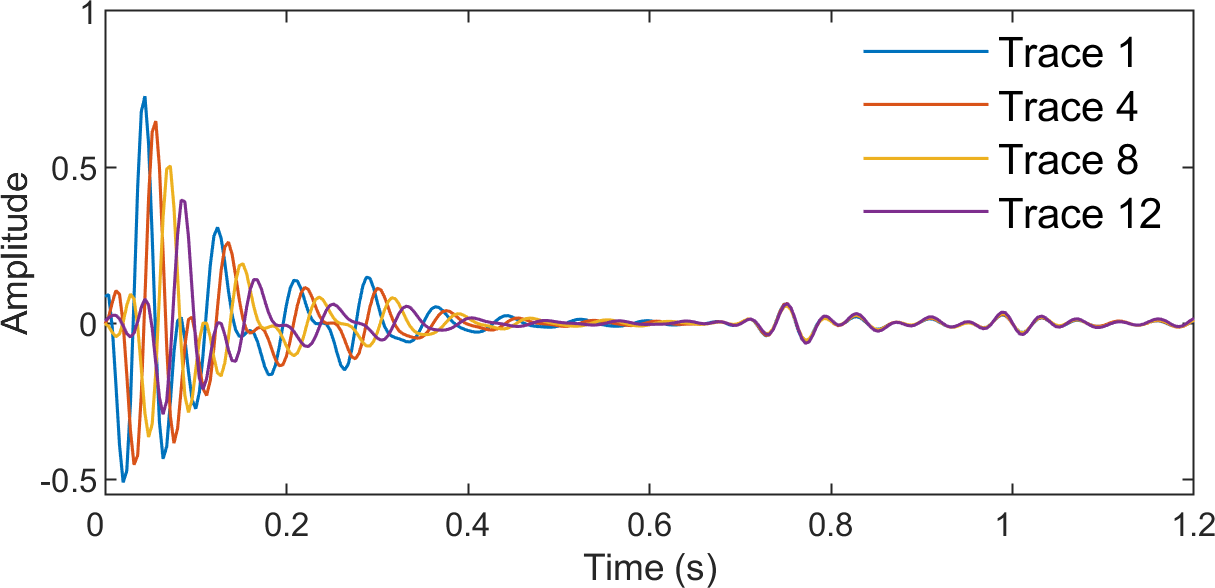}
\caption{Waveform comparison for the 1st, 4th, 8th, and 12th reconstructed near-offset traces in field data I.}
\label{fig14}
\end{figure}

To examine the reconstructed waveforms in detail, Figure~\ref{fig14} compares the time series for the 1st, 4th, 8th, and 12th reconstructed traces. These four traces span the entire missing zone from zero offset (trace 1) to the boundary with recorded data (trace 12). We can see that the waveforms exhibit progressive amplitude and phase variations across the four traces, reflecting realistic AVO behavior. Notably, the direct arrival and shallow reflection events show systematic amplitude decay and phase shifts as offset increases from trace 1 to trace 12. This AVO-consistent behavior is remarkable given that the model was trained exclusively on data starting from 150~m offset and never observed any near-offset traces during training. The progressive waveform variations indicate that the diffusion model has not simply copied or shifted recorded near-offset events into the missing zone, but has genuinely extrapolated dynamic wavefield attributes by propagating the AVO priors embedded in the far-offset training data.

\subsection{Application to field data II}

Field dataset II (Mobil AVO viking graben line 12) lacks approximately 250~m of near-offset data, equivalent to 10 missing traces at 25~m spacing. Thus, here we need to reconstruct the actual 0-250~m near-offset gap (10 missing traces) inherent to the acquisition. The model is retrained using the complete observed data starting from 250~m offset, following the training configuration established in validation experiment. 

\begin{figure}[htbp]
\centering
\includegraphics[width=\textwidth]{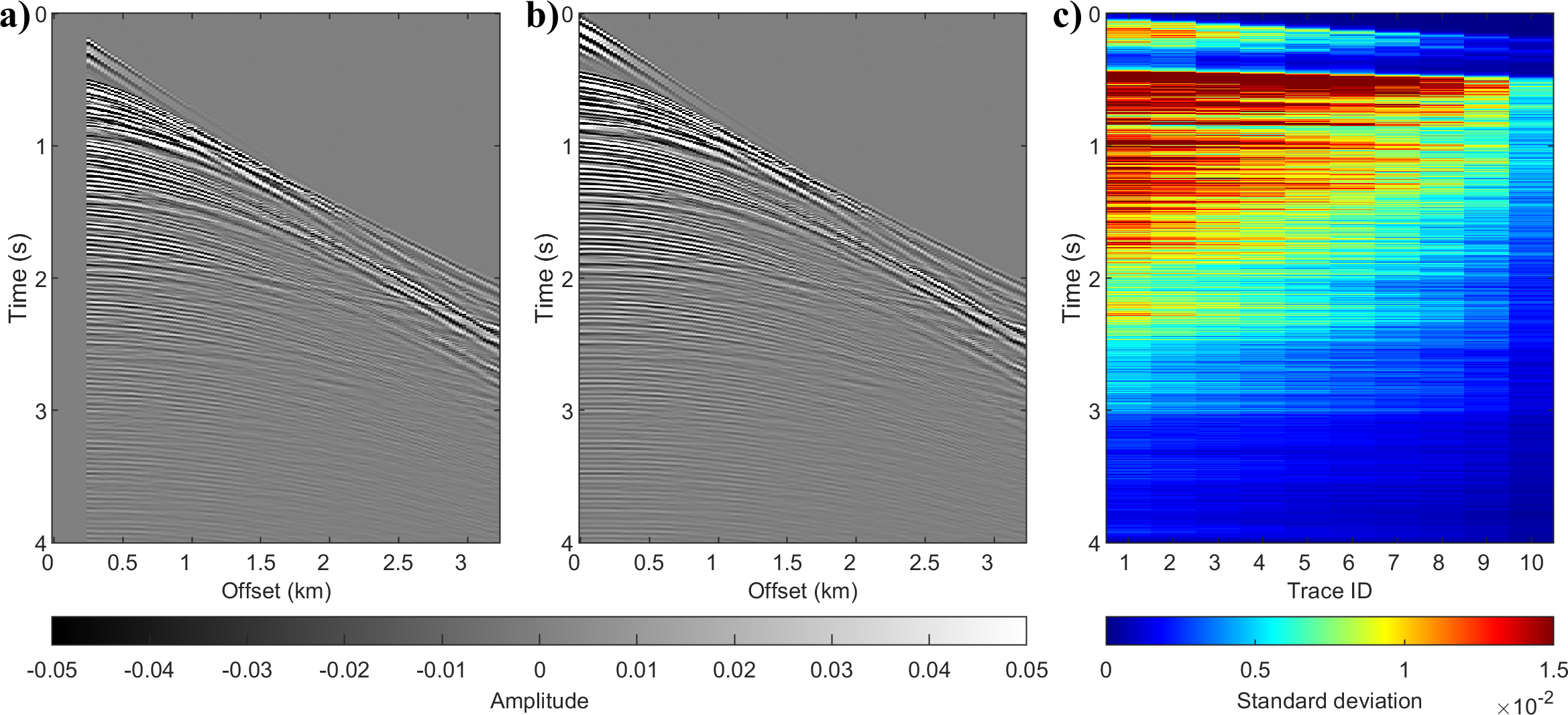}
\caption{Real near-offset gap reconstruction for field dataset II (Mobil AVO viking graben line 12). (a) Test shot gather displaying the complete offset range 0-3.25~km; recorded data begin at $\sim$0.25~km (trace 11). (b) Reconstructed complete section including the 10 newly generated near-offset traces (0-250~m). (c) Uncertainty map (standard deviation across 50 realizations) for the 10 reconstructed traces.}
\label{fig15}
\end{figure}

Figure~\ref{fig15} presents the reconstruction results for the actual 0-250~m gap. Panel (a) shows an example shot gather spanning 0-3.25~km offset, where recorded data begin at 0.25~km (trace 11). Panel (b) displays the reconstructed complete section, and panel (c) presents the uncertainty map across the 10 reconstructed traces. Despite the challenging spatial aliasing, the reconstruction achieves reasonable event continuity across the boundary between reconstructed and recorded traces. The reconstructed near-offset events maintain coherent moveout patterns. Notably, the direct arrival at zero offset again emerges at approximately zero time, demonstrating kinematic consistency. The uncertainty map (panel c) reveals substantially elevated uncertainty compared to field dataset I, consistent with the validation experiment. Meanwhile, the uncertainty distribution exhibits two clear trends:
\begin{itemize}
    \item First, in the offset dimension, uncertainty increases progressively from trace 10 (closest to recorded data at 250~m) toward trace 1 (zero offset), reflecting the fundamental challenge that extrapolation confidence degrades with increasing distance from conditioning information. 
    \item Second, in the time dimension, uncertainty displays a pronounced pattern: shallow events ($<2$~s) exhibit substantially higher uncertainty than deeper events ($>2$~s). 
\end{itemize}

This shallow-deep uncertainty contrast arises from the interplay of two factors. First, shallow reflections are subject to more severe spatial aliasing because their steeper moveout slope, characterized by rapid traveltime changes across the coarse 25~m receiver spacing. In contrast, deeper events exhibit gentler curvature and are consequently less aliased, allowing the model to establish more reliable statistical patterns during training. Simultaneously, shallow reflections typically have larger curvature near offset compared to deeper events. Higher curvatures inherently pose a greater challenege to extrapolation efforts of wavefields. Consequently, traces 1-5 at times $<2$~s display the highest uncertainty, where all adverse factors converge, while deeper portions of the reconstructed section show acceptably low uncertainty. This honest assessment of reconstruction limitations demonstrates the reliability of the uncertainty quantification framework, where the elevated uncertainty accurately reflects the genuine difficulty of extrapolating coherent information under severe aliasing and high-amplitude conditions.

\begin{figure}[htbp]
\centering
\includegraphics[width=0.75\textwidth]{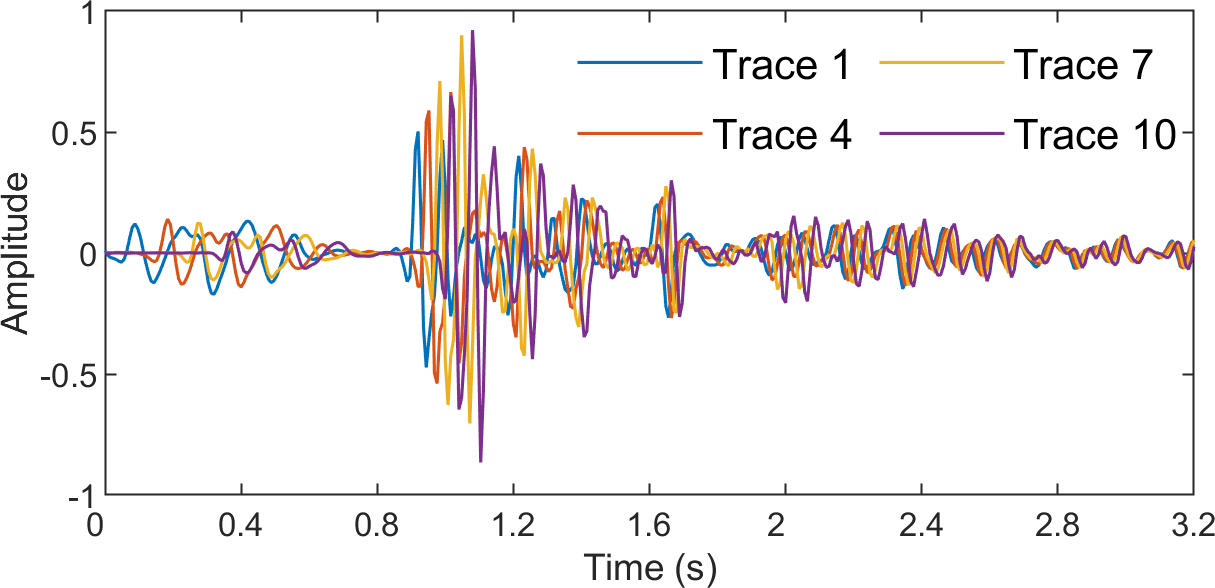}
\caption{Waveform comparison for the 1st, 4th, 7th, and 10th reconstructed near-offset traces in field data II.}
\label{fig16}
\end{figure}

Waveform analysis in Figure~\ref{fig16} compares the 1st, 4th, 7th, and 10th reconstructed traces spanning the 0-250~m missing zone. Similar to field dataset I (Figure~\ref{fig14}), the reconstructed waveforms exhibit clear AVO effects with progressive amplitude variations across traces, demonstrating that the diffusion model has successfully extrapolated offset-dependent amplitude behavior despite training only on far-offset data ($>250$~m). The four traces also display appropriate phase shifts across offsets, reflecting realistic moveout variations. Notably, the shallow strong reflection event below 2~s shows more pronounced phase shifts compared to deeper events, which is consistent with the steeper moveout curvature in the shallow zone that leads to more severe spatial aliasing, as reflected in the elevated uncertainty observed in Figure~\ref{fig15}c.

\section{\textbf{Discussion}}
The validation and practical application experimental results demonstrated that the proposed self-supervised diffusion framework achieves superior near-offset reconstruction performance compared to the conventional parabolic Radon transform across synthetic and field datasets. The quantitative metrics in Table~\ref{tab:metric_comparison} reveal substantial improvements in MAE, SSIM, and SNR, especially on well-sampled data. These performance enhancements stem from the method's ability to learn complex, offset-dependent statistical patterns directly from the available far-offset data and propagate this learned prior information toward near offsets through the recursive inference strategy. Importantly, the successful operational deployment on actual near-offset gaps validates the method's practical viability for real-world seismic processing workflows where ground-truth validation is impossible.

A particularly noteworthy capability of the proposed method is its effectiveness in extrapolating near-offset traces despite never having access to complete near-offset data during training. These exciting results, where we reconstruct information from a region entirely absent in the training set, are enabled by the diffusion model's capacity to capture and generalize the underlying wavefield physics embedded in the far-offset observations. Close examination of the waveform comparisons in both validation experiments (Figures~\ref{fig3}, \ref{fig6}, and \ref{fig9}) and real-world applications (Figures~\ref{fig13} and \ref{fig15}) reveals an even more remarkable phenomenon: the reconstructed near-offset traces exhibit amplitude variations across different offsets from what the diffusion model learned over many shots from the available data that are consistent with amplitude-versus-offset (AVO) effects. Specifically, the reconstructed waveforms display progressively varying amplitudes for the same reflection event, possibly mirroring the true AVO behavior caused by interface reflectivity changes with incident angle. This AVO preservation is particularly evident in the actual gap reconstructions, where no ground-truth near-offset data existed during either training or evaluation, yet the extrapolated traces still capture realistic offset-dependent amplitude trends. This suggests that the model has implicitly learned to extrapolate not only kinematic attributes (traveltimes and moveout curvature) but also dynamic attributes (AVO trends) from the far-offset training data.

The mechanism underlying this AVO-aware reconstruction can be understood through two complementary perspectives. First, the far-offset data inherently contain AVO information, as reflection amplitudes vary systematically across the recorded offset range due to angle-dependent reflectivity. By training on overlapping patches that span multiple offsets, the conditional diffusion model learns the statistical dependencies governing how amplitudes evolve as a function of offset. Crucially, training is performed across all shot gathers in the dataset simultaneously. This cross-shot training strategy is particularly powerful because the AVO information for a reflection event in one shot's missing near-offset zone may already be captured in another shot's recorded far-offset data due to different source-receiver geometries. For instance, a reflection at 100~m offset in shot A (missing) might correspond to the same geological interface at 300~m offset in shot B (recorded), allowing the model to learn that interface's amplitude behavior from shot B and transfer it to reconstruct shot A. Second, the recursive inference strategy plays a crucial role in propagating this learned prior from far to near offsets. At each step of the trace-by-trace reconstruction, the model conditions on the most recently available (reconstructed or recorded) offset information, effectively transferring amplitude trends gradually toward zero offset. This gradual extrapolation allows the model to maintain physically plausible amplitude behavior even in the absence of direct near-offset supervision.

The mechanism underlying this AVO-aware reconstruction can be understood through two complementary perspectives. First, the far-offset data inherently contain AVO information, as reflection amplitudes vary systematically across the recorded offset range due to angle-dependent reflectivity. By training on overlapping patches that span multiple offsets, the conditional diffusion model learns the statistical dependencies governing how amplitudes evolve as a function of offset. Second, the recursive inference strategy plays a crucial role in propagating this learned prior from far to near offsets. At each step of the trace-by-trace reconstruction, the model conditions on the most recently available (reconstructed or recorded) offset information, effectively transferring amplitude trends gradually toward zero offset. This gradual extrapolation allows the model to maintain physically plausible amplitude behavior even in the absence of direct near-offset supervision.

Beyond reconstruction accuracy, the proposed method offers a critical advantage through its natural uncertainty quantification capability. The ensemble sampling analysis in validation experiments (Figures~\ref{fig10}, \ref{fig11}, and \ref{fig12}) and real-world applications (Figures~\ref{fig13}c and \ref{fig15}c) reveals that the uncertainty maps consistently correlate with reconstruction error magnitude (when ground truth is available) and exhibit progressive increase toward smaller offsets, confirming that epistemic uncertainty genuinely reflects the difficulty of extrapolation at increasing distances from conditioning data. This self-assessment capability addresses a fundamental limitation in realistic near-offset reconstruction scenarios: the absence of ground-truth data for validation. In real-world seismic processing, geophysicists cannot directly evaluate reconstruction quality because complete near-offset traces are inherently unavailable, making the uncertainty maps an indispensable alternative quality indicator for identifying high-confidence regions suitable for subsequent processing and flagging low-confidence zones requiring manual inspection or alternative strategies.

However, the effectiveness of this approach is fundamentally tied to the spatial continuity and coherence of seismic events in the data. The proposed method relies on the assumption that neighboring offset traces exhibit smooth, predictable variations in both kinematics and dynamics. When this assumption holds, like in the synthetic dataset and field dataset I, the diffusion model successfully captures the underlying wavefield structure and achieves high-fidelity reconstruction. In contrast, when spatial continuity is degraded by factors such as coarse receiver spacing and pronounced spatial aliasing (as in field dataset II), the model's performance declines. The validation results show negative SNR values and elevated reconstruction errors (Table~\ref{tab:metric_comparison}), while the actual 0-260~m gap reconstruction exhibits substantially higher uncertainty, particularly in the shallow zone where steep moveout curvature exacerbates aliasing effects. These results reflect the inherent difficulty of extrapolating near-offset information from aliased far-offset observations, where high-frequency components are undersampled and event coherence is disrupted.

For datasets exhibiting severe spatial aliasing, a potential mitigation strategy is to apply dealiasing preprocessing prior to near-offset reconstruction. Techniques such as prediction-error filter \citep{chen20215d}, sparse mathematical transform \citep{schonewille2009seismic, naghizadeh2013multidimensional}, or deep learning-based dealiasing \citep{fang2021dealiased, wei2021aliased} could enhance spatial continuity by interpolating additional pseudo-traces or suppressing aliasing artifacts, thereby providing a more favorable input for the diffusion-based extrapolation. This direction represents a promising avenue for extending the proposed framework to more challenging acquisition scenarios.

Another practical consideration revealed by the field data I experiments is the importance of restricting training patch extraction to an appropriate offset range. As noted in field data I, we limited patch sampling to the first 200 channels rather than utilizing the entire offset aperture. This restriction addresses the amplitude scaling issue: far-offset traces often exhibit significantly larger amplitude variations due to geometric spreading, attenuation, and complex multiple interference, which can dominate the training loss and hinder the model's ability to learn the subtle amplitude trends relevant to near-offset reconstruction. By focusing the training on the near-to-mid offset region, the model learns patterns that are more representative of the target reconstruction zone, leading to improved generalization. This design choice highlights the importance of domain-specific considerations in adapting diffusion models to geophysical applications.

\section{\textbf{Conclusions}}
We developed a self-supervised diffusion-based framework for reconstructing missing near-offset traces in marine towed-streamer seismic data without requiring complete ground-truth training data. By leveraging overlapping patch extraction with single-trace shifts from the available far-offset section, the proposed method trains a conditional diffusion model to learn offset-dependent statistical patterns and performs trace-by-trace recursive inference from far to near offsets. Comprehensive validation experiments on synthetic and field datasets demonstrated superior performance compared to the conventional parabolic Radon transform. Operational deployment on actual near-offset gaps, where we reconstruct the inherent 0-150~m gap in field dataset I and the 0-260~m gap in field dataset II, further demonstrates the method's practical viability in real-world scenarios. Remarkably, both the validation experiments and real-world applications reveal that the reconstructed near-offset traces exhibit amplitude-versus-offset variations consistent with realistic offset-dependent amplitude behavior, despite the model never having access to near-offset data during training. This indicates successful extrapolation of both kinematic and dynamic wavefield attributes from the far-offset training data alone. Furthermore, the probabilistic nature of the diffusion model enables natural uncertainty quantification through ensemble sampling, providing practitioners with self-assessment capability to evaluate reconstruction reliability without requiring ground-truth near-offset references. While performance degrades on severely aliased data where spatial continuity is disrupted, the method demonstrated robust generalization across diverse acquisition scenarios and, therefore, offers a practical solution for operational seismic processing workflows.
\section{\textbf{Acknowledgment}}
This publication is based on work supported by the King Abdullah University of Science and Technology (KAUST). The authors thank the DeepWave sponsors for their support. The authors gratefully acknowledge CGG for providing the marine towed-streamer dataset acquired offshore northwest Australia. The authors also thank the Society of Exploration Geophysicists (SEG) for making the Mobil AVO Viking Graben Line 12 field dataset and the 2D SEG Advanced Modeling (SEAM) velocity model publicly available. This work utilized the resources of the Supercomputing Laboratory at King Abdullah University of Science and Technology (KAUST) in Thuwal, Saudi Arabia.
\vspace{0.5cm}
\section{\textbf{Code Availability}}
The data and accompanying codes that support the findings of this study are available at: 
\\
\url{https://github.com/DeepWave-KAUST/SSLDiff-Interpolation-pub}. (During the review process, the repository is private. Once the manuscript is accepted, we will make it public.)
\vspace{0.5cm}

\bibliographystyle{unsrtnat}
\bibliography{references}

\end{document}